\documentclass[structabstract]{aa}  
\usepackage{graphicx} 
\usepackage{txfonts}
\usepackage{natbib}
\usepackage{hyperref}
\usepackage{siunitx}  
\usepackage{xcolor}   

\newcommand{\HII}{\mbox{H\,{\sc ii}}}
\newcommand{\HeI}[1]{\mbox{He\,{\sc i}~$\lambda${#1}}}
\newcommand{\HeII}[1]{\mbox{He\,{\sc ii}~$\lambda${#1}}}

\newcommand{\NaId}[1]{\mbox{Na\,{\sc i}~$\lambda\lambda${#1}}}
\newcommand{\GG}{\mbox{$G$}}

\newcommand{\GGc}{\mbox{$G^\prime$}}

\newcommand{\GBPmGRPc}{\mbox{$G_{\rm BP}^\prime-G_{\rm RP}^\prime$}}

\newcommand{\AG}{\mbox{$A_G$}}
\newcommand{\Gabs}{\mbox{$G_{\rm abs}$}}
\newcommand{\chir}{\mbox{$\chi^2_{\rm red}$}}
\newcommand{\phz}{\phantom{0}}

\newcommand{\mci}[1]{\multicolumn{1}{c}{#1}}
\newcommand{\mcii}[1]{\multicolumn{2}{c}{#1}}

\newcommand{\pmra}{\mbox{$\mu_{\alpha *}$}}
\newcommand{\pmdec}{\mbox{$\mu_{\delta}$}}
\newcommand{\pmrag}{\mbox{$\mu_{\alpha *,{\rm g}}$ }}
\newcommand{\pmdecg}{\mbox{$\mu_{\delta,{\rm g}}$}}
\newcommand{\pmrac}{\mbox{$\mu^{\prime}_{\alpha *}$}}
\newcommand{\pmdecc}{\mbox{$\mu^{\prime}_{\delta}$}}

\newcommand{\Teff}{\mbox{$T_{\rm eff}$}}
\newcommand{\mi}{\mbox{$m_{\rm i}$}}

\newcommand{\EBV}{\mbox{$E(4405-5495)$}}
\newcommand{\RV}{\mbox{$R_{5495}$}}

\newcommand{\VO}[1]{Villafranca~O-{#1}}

\newcommand{\lili}{LiLiMaRlin}
\newcommand{\tf}{\mbox{$t_{\rm f}$}}
\newcommand{\dmin}{\mbox{$d_{\rm min}$}}
\newcommand{\alphamin}{\mbox{$\alpha_{\rm min}$}}
\newcommand{\deltamin}{\mbox{$\delta_{\rm min}$}}
\newcommand{\tmin}{\mbox{$t_{\rm min}$}}

\begin{document}

   \title{Escape from the Bermuda cluster: \linebreak
          orphanization by multiple stellar ejections}
   \titlerunning{The origin of the orphan cluster in the North America nebula}

   \author{J. Ma\'{\i}z Apell\'aniz \inst{1}
           \and
           M. Pantaleoni Gonz\'alez\inst{1,2}
           \and
           R. H. Barb\'a\inst{3}
           \and
           M. Weiler\inst{4}
           }
   \authorrunning{J. Ma\'{\i}z Apell\'aniz et al.}

   \institute{Centro de Astrobiolog\'{\i}a, CSIC-INTA. Campus ESAC. 
              C. bajo del castillo s/n. 
              E-\num{28692} Villanueva de la Ca\~nada, Madrid, Spain.\linebreak
              \email{jmaiz@cab.inta-csic.es} 
              \and
              Departamento de Astrof{\'\i}sica y F{\'\i}sica de la Atm\'osfera, Universidad Complutense de Madrid. 
              E-\num{28040} Madrid, Spain. 
              \and
              Departamento de Astronom{\'\i}a, Universidad de La Serena.
              Av. Cisternas 1200 Norte.
              La Serena, Chile.
              \and
              Departament de F{\'\i}sica Qu\`antica i Astrof{\'\i}sica, Institut de Ci\`encies del Cosmos (ICCUB). Universitat de Barcelona (IEEC-UB). 
              Mart{\'\i} i Franqu\`es 1.
              E-\num[minimum-integer-digits=5]{08028} Barcelona, Spain. \\
             }

   \date{Received 4 October 2021 / Accepted 12 November 2021}

 
  \abstract
   {Dynamical interactions in young stellar clusters can eject massive stars early in their lives and significantly alter their
    mass functions. If all of the most massive stars are lost, we are left with an orphan cluster.}
   {We study the Bermuda cluster (\VO{014}~NW), the most significant young stellar group in the North America and Pelican nebulae, and the massive 
    stars that may have been ejected from it to test if it has been orphaned.}
   {We use {\it Gaia}~EDR3 parallaxes and proper motions to search for walkaway and runaway stars in the vicinity of the North America
    and Pelican nebulae. The candidates are analyzed with a combination of spectroscopy and photometry to assess their nature and their
    trajectories are traced back in time to determine at what time they left the Bermuda cluster.}
   {We detect three ejection events, dubbed the Bajamar, Toronto, and HD~\num{201795} events, that expelled (a minimum of) 5, 2, and 2 systems, respectively, or 
    6, 3, and 3 stars if we count the individual components in spectroscopic/eclipsing binaries. The events took place 1.611$\pm$0.011~Ma, 1.496$\pm$0.044~Ma, and 
    1.905$\pm$0.037~Ma ago, respectively, but our analysis is marginally consistent with the first two being simultaneous. 
    We detect bow shocks in \textit{WISE} images associated with four of the ejected systems and their orientation agrees with that of 
    their relative proper motions with respect to the cluster. Combining the three events, the Bermuda cluster has lost over 200~M$_\odot$, including its three most 
    massive stars, so it can be rightfully considered an orphan cluster. One consequence is that the present-day mass function of the cluster has been radically altered 
    from its top-heavy initial value to one compatible with a Kroupa-like function. Another one is that the cluster is currently expanding with a dynamical time scale 
    consistent with the cause being the ejection events. A scenario in which the Bermuda cluster was formed in a conveyor belt fashion over several hundreds of ka or even 
    1~Ma is consistent with all the observables.}
   {}
   \maketitle
%

\section{Introduction}

$\,\!$\indent Dynamical interactions between three or more bodies in stellar clusters and supernova explosions in binary systems
can lead to stars moving at high speed through the Galaxy \citep{Zwic57,Blaa61,Poveetal67}. The ejected stars with velocities higher
than 30 km/s are called runaway stars \citep{Hoogetal01} and can be easily found with {\it Gaia} astrometry \citep{Maizetal18b} while
the slower ones, called walkaway stars, are more common but more difficult to identify \citep{Renzetal19b}. Ejections favor 
high-mass stars over low-mass ones and tend to occur early in the cluster life \citep{OhKrou16}. Hence, such events can
have a large impact on the massive-star mass function \citep{PflaKrou06} and, in extreme cases, they can produce ``orphan clusters'' 
where the most massive stars have disappeared from the cluster altogether and the present-day mass function (PDMF) is capped with respect 
to the initial one or IMF (\citealt{Ohetal15}; Villafranca~II, 
{\citealt{Maizetal21e}).} 

\begin{table*}
\caption{Basic data for the sample of runaway/walkaway stellar systems ejected from the Bermuda cluster analyzed in this paper.}
\centerline{
\begin{tabular}{llclll}
\hline
Star                      & {\it Gaia} EDR3 ID        & Event           & Spectral type      & Ref. & Notes                                \\
\hline
Bajamar star              & \num{2162889493831375488} & Bajamar         & O3.5 III(f*) + O8: & M16  & SB2 orbit analysis in progress       \\
Tyc~3179-00756-1          & \num{2162884099352103936} & Bajamar         & ---                & ---  & Late-O/early-B from CHORIZOS         \\
HD~\num{195965}           & \num{2083681294654108672} & Bajamar         & B0.2 Vn            & ViII & ---                                  \\
Tyc~3575-01514-1          & \num{2163288685285250304} & Bajamar         & ---                & ---  & Late-B from CHORIZOS                 \\
Tyc~3157-00918-1          & \num{2067356398827733504} & Bajamar         & ---                & ---  & Late-B/early-A from CHORIZOS         \\
\hline
Toronto star              & \num{2163075895429683072} & Toronto         & O6.5 V((f))z       & S11  & B companion, \citet{Willetal01}      \\
HDE~\num{227090}          & \num{2060123330345685248} & Toronto         & ---                & ---  & Bad RUWE, mid/late-B from CHORIZOS   \\
\hline
HD~\num{201795}           & \num{1968091458287111296} & HD~\num{201795} & B0.5 V             & ViII & ---                                  \\
HD~\num{200776}           & \num{2163574661395444864} & HD~\num{201795} & B1 IV:p            & M55  & Eclipsing binary, \citet{Dervetal11} \\
\hline
\multicolumn{6}{l}{References. M55: \citet{Morgetal55}. M16: \citet{Maizetal16}. S11: \citet{Sotaetal11a}. ViII: Villafranca~II.}
\end{tabular}
}
\label{sample}
\end{table*}

In the Villafranca project we are combining astrometry and photometry from {\it Gaia} \citep{Prusetal16} and spectroscopy from the 
Galactic O-Star Spectroscopic Survey (GOSSS, \citealt{Maizetal11}) and \lili, \citep{Maizetal19a} to study the properties of Galactic
stellar groups with OB stars. In Villafranca~I \citep{Maizetal20b} we analyzed 16 OB groups using {\it Gaia}~DR2 data and in Villafranca~II 
{\citep{Maizetal21e}} 
we added another ten and reanalyzed the 26 objects with {\it Gaia}~EDR3 data. In Villafranca~I we 
discussed the case of \VO{014}, the OB group associated with the North America and Pelican nebulae. {\it Gaia}~DR2 data confirmed that 
the Bajamar star (named after its ``geographical'' location with respect to the North America nebula at the Bahamas), an early-type~O SB2
system previously suspected to be the main ionizing source of the nebulae \citep{ComePasq05,Maizetal16}, is indeed at a similar distance 
as the nebula but at a location where no cluster is present (see \citealt{KuhnHill20} and Villafranca~II for confirmation of the distance 
with {\it Gaia} EDR3). \citet{Kuhnetal20} showed that the Bajamar star is a walkaway stellar system recently ejected from \VO{014}~NW, the most 
significant cluster in the region and located around Bermuda (following the geographical nomenclature). The other previously known O-type
system in the vicinity, HD~\num{199579} (which can be dubbed the Toronto star following the geographical nomenclature), was apparently
beyond the nebula using {\it Gaia}~DR2 data (\citealt{Kuhnetal20} and Villafranca~I) but {\it Gaia}~EDR3 brings it within its reach 
(Villafranca~II) when parallaxes are recalibrated according to 
{\citet{Maiz21}.} 
The Toronto star has a proper motion consistent with also being ejected from the Bermuda cluster as a walkaway system \citep{Kuhnetal20}.

{The North America and Pelican nebulae constitute a complex star-forming region whose pareidolic common names arise from the molecular 
clouds in the foreground of the \HII\ region \citep{BallScov80,Zhanetal14,Kongetal21}. As discussed below, massive-star formation goes back at least 2~Ma but the
region is still currently forming new stars, as evidenced by the study of outflows and sub-mm cores of \citet{Balletal14}.} 

In this paper we present a scenario in which the Bajamar and Toronto stars (among other high- and intermediate-mass stars) were ejected 
soon after the formation of the Bermuda cluster.

\section{Data and methods}

$\,\!$\indent In previous papers (\citealt{Maiz19}, Villafranca~I) we developed a method that uses astrometric and photometric information from {\it Gaia} to
determine the distances and proper motions of OB groups. In Villafranca~II we applied the method to the Bermuda cluster using {\it Gaia}~EDR3 data and derived
a distance $d$ of $798\pm 6$~pc, a cluster center at $\alpha = 313.10^{\rm o}$ and $\delta = +44.40^{\rm o}$ (J2000), 
and a group proper motion \pmrag\ of $-1.370\pm 0.020$~mas/a and \pmdecg\ of $-3.054\pm 0.020$~mas/a. Those values use the 
{\it Gaia}~EDR3 astrometric calibration of 
{\citet{Maiz21}} 
(see also \citealt{Lindetal21a,Lindetal21b,Maizetal21c}) and they were
calculated with the correction for proper motions of \citet{CanGBran21}. The calibration and the correction are also applied to the rest of the data in this paper.

In this paper we search for stellar systems that could have been ejected from the Bermuda cluster. For that purpose, we downloaded all the {\it Gaia}~EDR3 sources within 
15 degrees of the cluster and with $G < 14$~mag. For each object we applied the calibration described in the previous paragraph and computed the individual
distances (see Appendix~A) and relative proper motions with respect to the cluster:

\begin{equation}
 \pmrac = \pmra - \pmrag, \; \pmdecc = \pmdec - \pmdecg.
 \label{pm}
\end{equation}

\noindent Based on those, we did an initial filtering of the {\it Gaia}~EDR3 sources whose relative proper motions point away from the Bermuda cluster (within a tolerance angle)
and whose corresponding (approximate) flight times \tf\ (as initially estimated from their separations with the cluster and relative proper motions) are less 
than 2.5~Ma. We already knew that this was the case for the Bajamar and Toronto stars (see above) and to those we added another three candidate early-B stars that 
passed the initial filtering (HD~\num{195965}, HD~\num{201795}, and HD~\num{200776}, Tables~\ref{sample}~and~\ref{indivresults}) 
and that were already included in the ALS catalog \citep{Pantetal21}. We then traced back the trajectories of those five systems to find out if any two of them 
had been at the same location at the same time within the last 2.5~Ma. Doing that properly requires tracing the 3-D trajectories of each pair of systems including 
[a] the effect of the Galactic gravitational potential and [b] the possible range of variations allowed by the uncertainties in proper motions (relatively small), 
distances (larger), and radial velocities $v_{\rm r}$ (a priori unknown). To trace back the trajectories in the Galaxy, we assumed a potential that can be separated
in $x+y$ (the Galactic mid-plane coordinates with $x$ pointing towards the Galactic Center and $y$ towards the rotation direction) and $z$ (the vertical Galactic 
direction towards the North). For the $x+y$ directions we used the potential associated with a circular velocity with an exponential dependence on 
$r = \sqrt{x^2+y^2}$ and a characteristic length scale of 1~kpc, adjusting the values to be consistent with the distance to the Galactic Center and rotation speed 
of the local standard of rest (LSR) of \citet{Abutetal19} and the epicycle frequency associated with the Oort constants of \citet{Lietal19}. For the vertical direction 
we used a potential derived from the mass density of \citet{Monietal12}. The velocity of the Sun with respect to the LSR was taken from \citet{Schoetal10b}.

\begin{table*}
\caption{Astrometric characteristics of the sample of runaway/walkaway stellar systems in this paper.} 
\centerline{
\begin{tabular}{llr@{$\pm$}lr@{$\pm$}lr@{$\pm$}lrr@{$\pm$}lr@{$\pm$}l}
\hline
Star                      & \mci{$d$}          & \mcii{\pmra}     & \mcii{\pmdec}   & \mcii{$v_{\rm t}$} & \mci{$m_{\rm i}$}  & \mcii{$m_{\rm i}v_\alpha$} & \mcii{$m_{\rm i}v_\delta$} \\
                          & \mci{(pc)}         & \mcii{(mas/a)}   & \mcii{(mas/a)}  & \mcii{(km/s)}      & \mci{(M$_\odot$)}  & \mcii{(M$_\odot$km/s)}     & \mcii{(M$_\odot$km/s)}     \\
\hline
Bajamar star              & $755^{+24}_{-22}$  &  $-$0.224&0.050  &  $-$4.554&0.053 &      7.1&0.1       & {           106.0} &  $+$650&19                 &   $-$371&42                \\ 
Tyc~3179-00756-1          & $798^{+13}_{-13}$  &  $-$0.654&0.029  &  $-$4.099&0.029 &      4.7&0.1       & {            18.4} &    $+$68&4                 &    $-$52&7                 \\ 
HD~\num{195965}           & $790^{+61}_{-53}$  &  $-$9.444&0.116  &  $+$5.311&0.103 &     43.0&0.8       & {            15.5} &   $-$421&12                &   $+$516&6                 \\ 
Tyc~3575-01514-1          & $791^{+16}_{-16}$  &  $-$0.432&0.035  &  $-$0.909&0.035 &      9.8&0.4       & {             3.6} &    $+$16&1                 &    $+$31&1                 \\ 
Tyc~3157-00918-1          & $750^{+10}_{-10}$  & $-$10.923&0.028  & $-$13.793&0.029 &     42.8&1.0       & {             3.6} &  $-$113&3                  &   $-$105&2                 \\ 
Toronto star              & $846^{+74}_{-63}$  &  $+$0.464&0.093  &  $-$2.092&0.106 &     10.4&0.6       & {            41.0} &  $+$347&18                 &   $+$247&22                \\ 
HDE~\num{227090}          & $700^{+140}_{-99}$ & $-$25.015&0.238  & $-$24.892&0.261 &    100.7&7.8       & {             4.5} &  $-$379&28                 &   $-$250&20                \\ 
\hline
HD~\num{201795}           & $796^{+45}_{-40}$  &  $+$4.402&0.064  & $-$13.272&0.071 &     46.0&3.3       & {            12.4} &   $+$321&11                &   $-$472&42                \\ 
HD~\num{200776}           & $865^{+47}_{-42}$  &  $+$2.358&0.071  &  $+$0.496&0.073 &     20.7&0.2       & {            16.3} &   $+$238&3                 &   $+$240&4                 \\ 
\hline
\multicolumn{13}{p{15cm}}{Notes: The first three columns with values are the input data for the trajectory 
         calculation and the last three are results derived from it at the moment of ejection in the frame of reference of the cluster: the velocity in the plane of the sky 
         $v_{\rm t}$ and the two components of the linear momentum in the plane of the sky. Stellar initial masses are taken from Table~\ref{chorizos1} accounting for the masses
         of Bajamar B, Toronto B, and HD~\num{200776}~B (see Appendix~A). The uncertainties in linear momentum do not include uncertainties in the masses but just the ones 
         associated with velocities. The velocities at the time of ejection do not include the deceleration caused by the escape from the stellar system or the cluster and only 
         include the (small) effect of the Galactic potential on the current proper motions.}
\\
\end{tabular}
}
\label{indivresults}
\end{table*}

As described below in each case, those initial five systems could be grouped into three different events with a relatively well determined \tf\ within that period frame,
two each in two events and a third one with no counterpart(s). As a second step in our search for systems possibly ejected from the Bermuda cluster, we returned to the full 
filtered sample to detect if those ejection events had included any additional {\it Gaia}~EDR3 sources. We selected the systems with distances similar to that of the Bermuda 
cluster and the appropriate \tf\ estimates and added each candidate one by one as an additional system in each ejection event to a genetic algorithm that searches through all 
trajectories compatible with the measured proper motions and distances (within two sigmas of each uncertainty) to find which objects could have been within 0.1~pc at the 
time of the event. A genetic algorithm was used at this stage because of the high-dimensionality of the problem and the need to perform tests for a large number of systems. This 
process yielded additional candidates that were further tested in more detail with an implementation of the TNMIN routine of \citet{Mark09} that uses the truncated Newton 
method to find the minimum of a function. In a first pass we did a detailed calculation that required at least one encounter with a minimum simultaneous 3-D separation \dmin\ 
between all systems involved in the event of less than 6~AU\footnote{Such a search is computationally expensive and subject to round-off errors. The limit was set at 6~AU 
but it could have been continued until an arbitrarily small separation was achieved. Also note that we are not including gravitational focusing in our calculations,
which should be significant for an event involving massive stars in increasing the likelihood of a small \dmin.}. 
That pass gave us our final sample of 5, 2, and 2 systems involved in each of the three 
ejection events, respectively. In a second pass we set up a more lenient minimum \dmin\ of 200~AU to speed up the convergence of the search (once again, neglecting
gravitational focusing) and ran Monte Carlo simulations changing the input parameters (with a two-sigma tolerance, as above) until we found 100 circumstances of each of
the three events. This allowed us to calculate the statistical properties (mean, standard deviation, and correlations) of each of the input parameters and of the
coordinates (\alphamin, \deltamin) and flight times (\tmin) of each event, either without weighting or using the weights from the input parameters assuming a Gaussian 
distribution for each one of them (Table~\ref{eventresults}).

To gain additional information about each of the nine systems involved in the three ejection events plus eight current members of the Bermuda cluster, we used the spectral 
types determined from the Galactic O-Star Spectroscopic Survey (GOSSS, \citet{Maizetal11}) and we derived their stellar parameters from their photometry using 
CHORIZOS \citep{Maiz04c}. Details are given in Appendix~A. We also analyzed {\it Wide-field Infrared Survey Explorer} ({\it WISE}, \citealt{Cutretal13}) images to search for
bow shocks associated with the ejected stars (Appendix~B) and light curves from the {\it Transiting Exoplanet Survey Satellite} ({\it TESS}, \citealt{Ricketal15}) for the nine 
systems to study their variability (Appendix~D).

\begin{table*}
 \caption{Characteristics of the three ejection events analyzed in this paper.} 
\centerline{
\begin{tabular}{lcccr@{$\pm$}lr@{$\pm$}lr@{$\pm$}lcccr@{$\pm$}lr@{$\pm$}l}
\hline
Event           & $N_\star$ & weighted & cluster    & \mcii{\tmin} & \mcii{\alphamin} & \mcii{\deltamin} & $r_{t\alpha}$ & $r_{t\delta}$ & $r_{\alpha\delta}$ \\
                &           &          & motion     & \mcii{(ka)}  & \mcii{(deg)}     & \mcii{(deg)}     &               &               &                    \\
\hline
Bajamar         & 5         & no       & kept       & 1611&12      & 313.575&0.011    & 44.099&0.020     & $-$0.54       & $-$0.69      & $+$0.45             \\
                &           & yes      & kept       & 1611&11      & 313.577&0.011    & 44.095&0.015     & $-$0.75       & $-$0.72      & $+$0.69             \\
                &           & no       & subtracted & \mcii{}      & 313.107&0.014    & 44.442&0.019     & $-$0.74       & $-$0.62      & $+$0.53             \\
                &           & yes      & subtracted & \mcii{}      & 313.109&0.014    & 44.440&0.014     & $-$0.86       & $-$0.65      & $+$0.69             \\
\hline
Toronto         & 2         & no       & kept       & 1503&51      & 313.403&0.045    & 44.045&0.051     & $-$0.31       & $-$0.95      & $+$0.41             \\
                &           & yes      & kept       & 1496&44      & 313.434&0.039    & 44.062&0.045     & $-$0.28       & $-$0.96      & $+$0.39             \\
                &           & no       & subtracted & \mcii{}      & 312.971&0.053    & 44.367&0.042     & $-$0.59       & $-$0.92      & $+$0.66             \\
                &           & yes      & subtracted & \mcii{}      & 313.005&0.045    & 44.382&0.037     & $-$0.57       & $-$0.94      & $+$0.65             \\
\hline
HD~\num{201795} & 2         & no       & kept       & 1923&62      & 313.654&0.047    & 43.947&0.070     & $-$0.47       & $-$0.93      & $+$0.18             \\
                &           & yes      & kept       & 1905&37      & 313.641&0.039    & 43.985&0.039     & $-$0.59       & $-$0.89      & $+$0.22             \\
                &           & no       & subtracted & \mcii{}      & 313.076&0.061    & 44.346&0.060     & $-$0.72       & $-$0.91      & $+$0.43             \\
                &           & yes      & subtracted & \mcii{}      & 313.070&0.048    & 44.381&0.036     & $-$0.76       & $-$0.85      & $+$0.36             \\
\hline
\multicolumn{17}{p{15cm}}{Notes: The columns give (a) the number of systems involved in each event; (b) whether the
         results are weighted or not by the Gaussian factors, (c) whether the cluster motion has been subtracted or not; and (d) the flight
         times, event positions, and their correlations.}
\\
\end{tabular}
}
\label{eventresults}
\end{table*}

\section{Results}

$\,\!$\indent The procedure described above yielded three ejection episodes, which we dub the Bajamar, Toronto, and HD~\num{201795} events, respectively, after the 
earliest system involved in each one of them (Table~\ref{eventresults}). 
They are described one by one in this section. Prior to that, we briefly analyze the motion of the (center of the) Bermuda 
cluster to clarify the different coordinate systems involved. We end the section with an analysis of the internal cluster motions.

\subsection{The motion of the cluster center}

$\,\!$\indent Before we analyze the motion of the individual systems, we discuss what we know about the motion of the Bermuda cluster itself. The values above for its distance 
and proper motion are quite precise, as the first is known to better than 1\% and the second has uncertainties of just 0.020~mas/a in each coordinate. However, its $v_{\rm r}$
is a priori unknown. Could it be a significantly large value, thus implying that its distance has changed substantially in the last few Ma? To assess that possibility, we 
calculate the relevant velocity vector that corresponds to the measured proper motions. First, we subtract the effect in the observed cluster proper motion of the velocity of the Sun 
with respect to the LSR at our current location. Then, we calculate the cluster velocity in the plane of the sky with respect to the Sun LSR along Galactic longitude and latitude. 
Finally, we subtract the effect of the difference between the LSR at the cluster location and at the Sun's location to obtain the velocity of the cluster with respect to its
LSR in the plane of the sky. It is moving at 1 km/s towards the Galactic West (decreasing $l$) and at 4 km/s towards the Galactic North (increasing $b$)
i.e. its peculiar motion in those two directions is small,
amounting to $\sim$8~pc in 2~Ma. Therefore, its peculiar motion in the radial direction is also likely to be of that order or lower. As 8~pc is the uncertainty in the cluster distance 
itself (and distance uncertainties for individual systems are larger), we can assume that in the past the cluster (and any events that took place in it) was at a distance similar to the
current one. From a point of view of trajectory modelling, we will initially estimate that the ejection events took place at the same distance as the current value for the cluster but
will leave the value itself as a free parameter for the algorithm to determine.

A knowledge of the past trajectory of the cluster is also useful for two reasons: to calculate the ejection velocities with respect to the cluster, as that is the relevant frame to 
evaluate whether the ejected systems are walkaways or runaways and to estimate the conservation of linear momentum (Table~\ref{indivresults}), and to calculate the position of each event 
with respect to the cluster center to assess whether all originate from the same exact site or not (Table~\ref{eventresults}).

\subsection{The Bajamar event}

\begin{figure*}
\centerline{\includegraphics[width=0.48\linewidth]{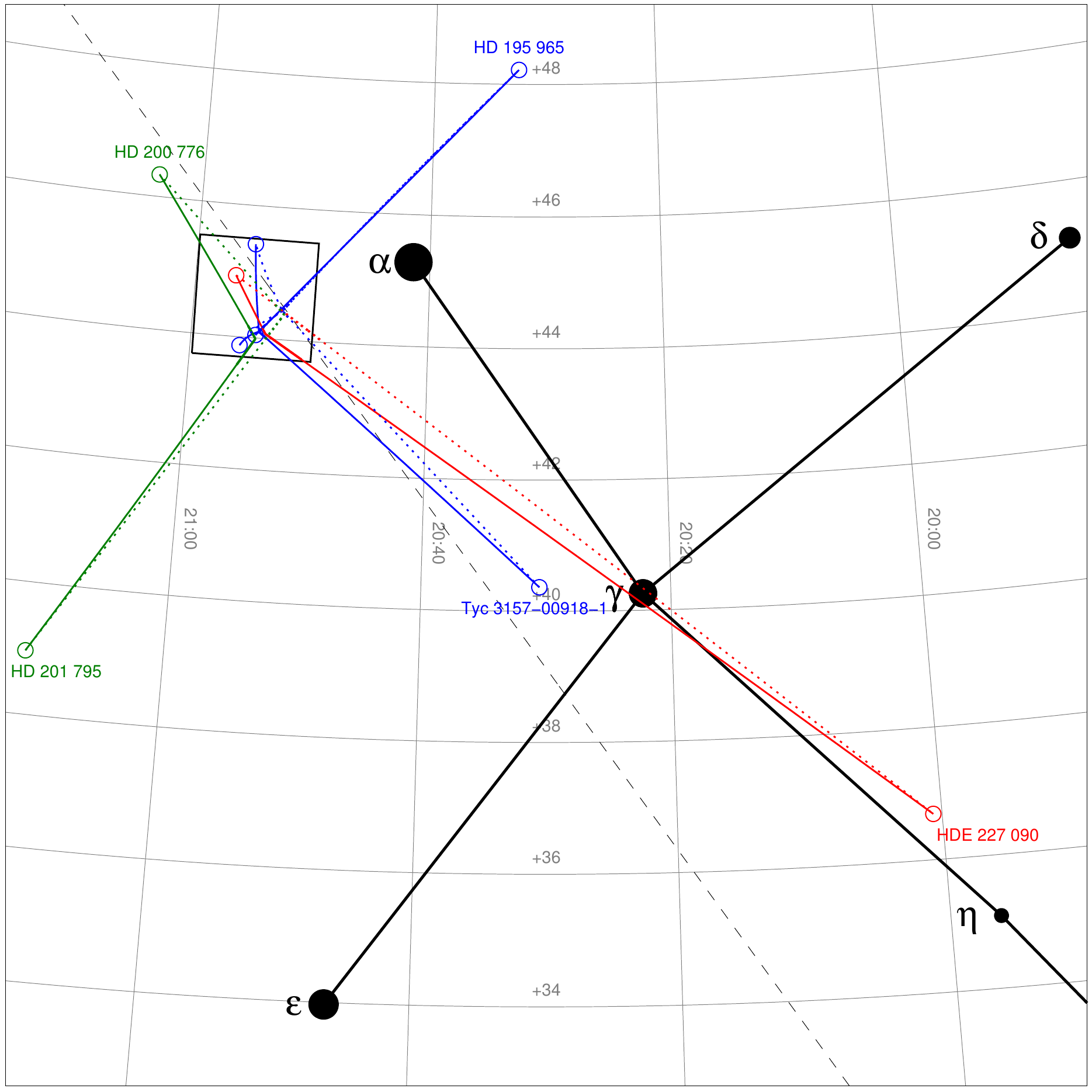} 
            \includegraphics[width=0.48\linewidth]{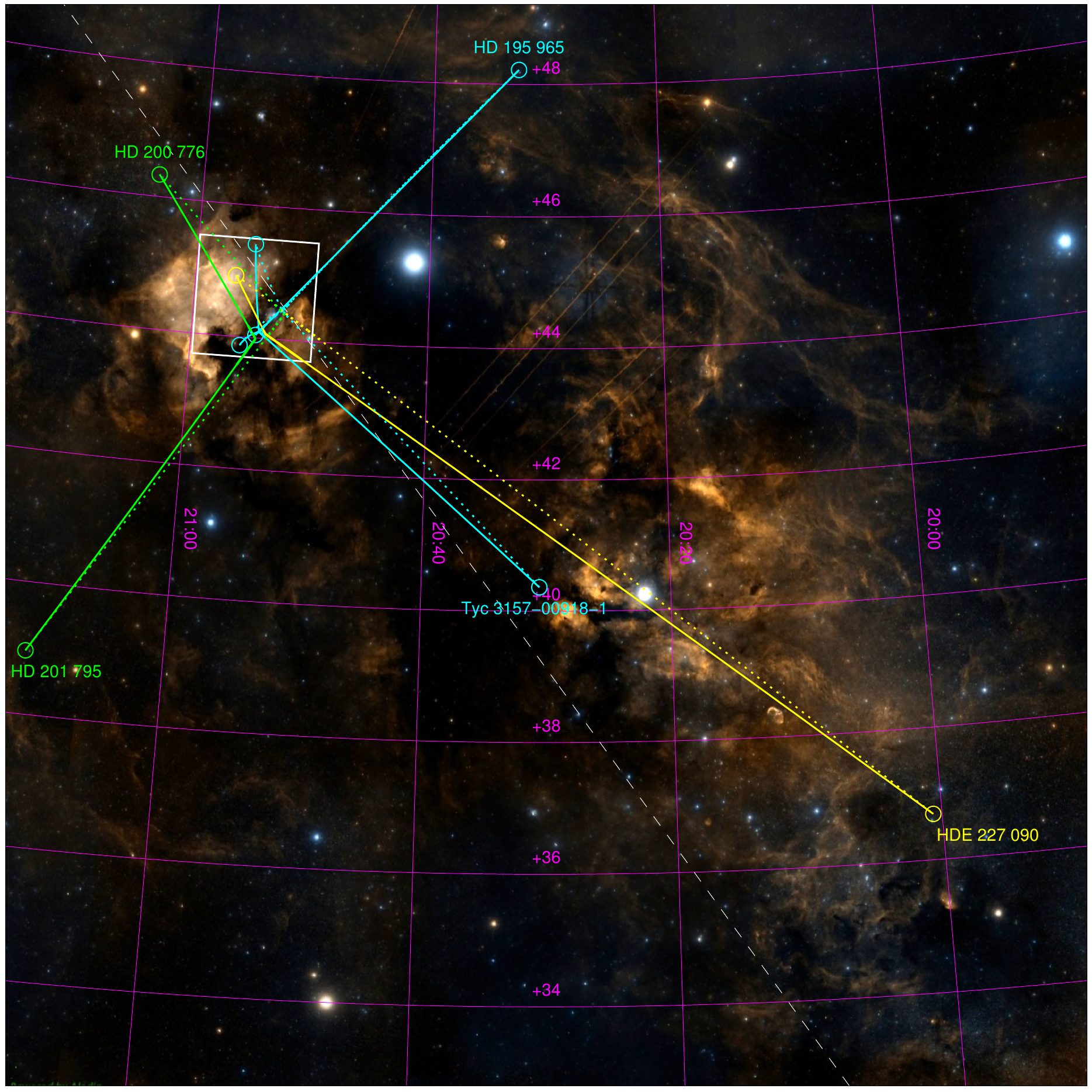}}
\centerline{\includegraphics[width=0.48\linewidth]{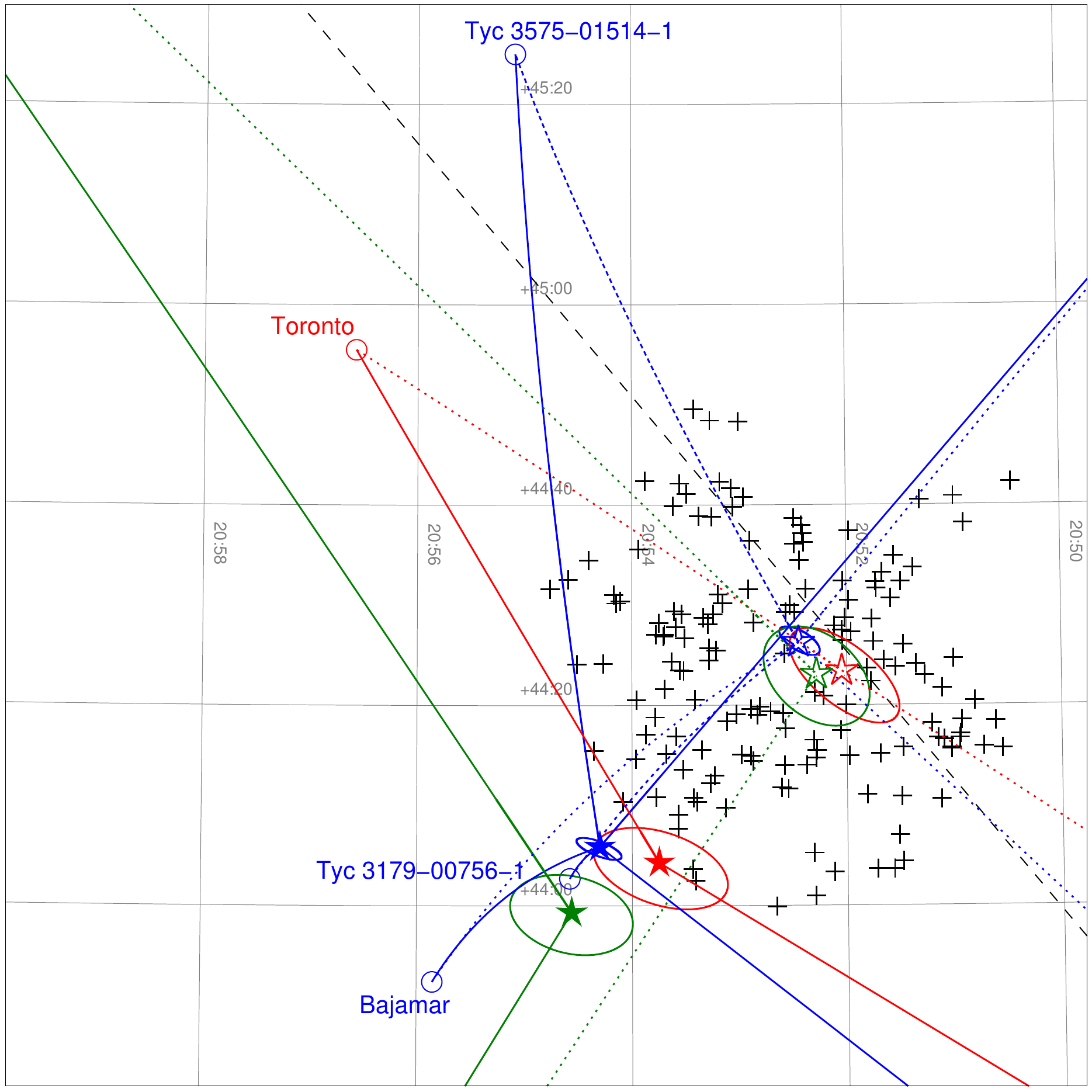} 
            \includegraphics[width=0.48\linewidth]{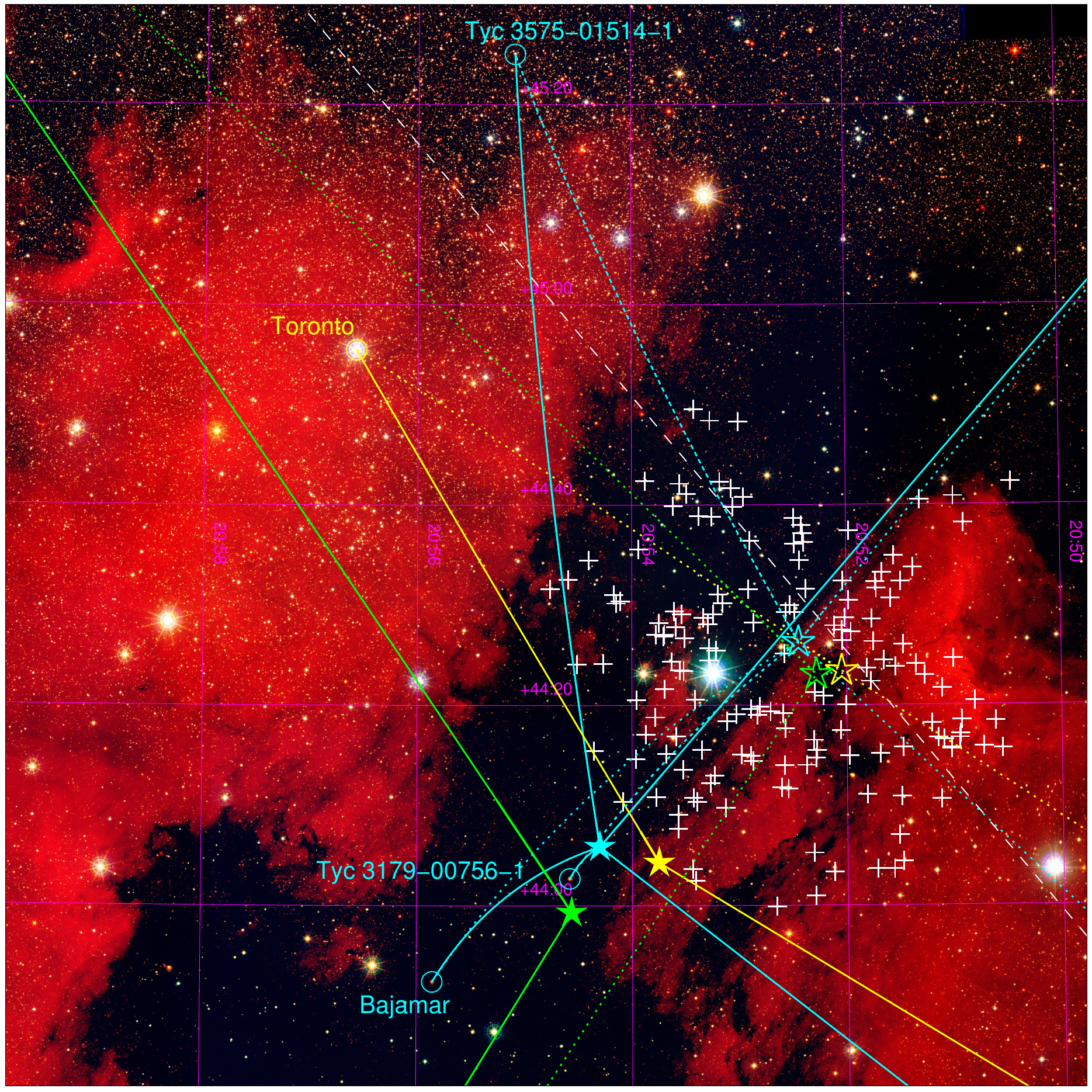}}
\caption{(top left) Chart in equatorial coordinates of the three ejection events in this paper, color-coded in blue (Bajamar), red (Toronto), and dark green (HD~\num{201795}).
         Colored solid lines show a representative trajectory for each system in the Sun's LSR and colored short-dashed lines the equivalent after subtracting the motion of 
         the Bermuda cluster. The black long-dashed line shows the Galactic equator and the square indicates the location of the bottom panels. The Cygnus asterism 
         is also plotted for reference. (top right) DSS2 image of the same field as the top left panel. Trajectories are color-coded here in cyan (Bajamar), yellow (Toronto), 
         and green (HD~\num{201795}). (bottom left) Equivalent chart of the square region in the top panels. Stars mark the location of each event in the Sun's LSR (filled) and 
         after subtracting the motion of the Bermuda cluster (non-filled). The ellipses mark the corresponding two-sigma uncertainty regions of each event using Gaussian weights.
         The black long-dashed line shows the Galactic equator and the black crosses mark the cluster 
         members from Villafranca~II. (bottom right) GALANTE \citep{Maizetal21d} three-color mosaic (red: F861M+F660N, green: F665N+F515N, blue: F450N+F420N) image of the 
         same field as the bottom left panel.
         {The slight curvature of the trajectories is caused by the Galactic potential.} 
         }
\label{chartimage}
\end{figure*}

$\,\!$\indent The Bajamar event involved at least five stellar systems and took place 1.611$\pm$0.011~Ma (with Gaussian weights) or 1.611$\pm$0.012~Ma ago (without Gaussian weights),
with a total possible flight-time range in the 100 Monte Carlo simulations of 1.586-1.644~Ma. The five stellar systems we detect with a common origin are the Bajamar star, Tyc~3179-00756-1, 
HD~\num{195695}, Tyc~3575-01514-1, and Tyc~3157-00918-1, with a total of at least six stars (including four massive stars) as the first one is a spectroscopic binary (Appendix~C). This 
event is the most significant one of the three detected due to its high number of stars involved, the large total amount of mass lost by the cluster in the episode, and the inclusion of 
Bajamar A (the earliest and most massive star in the region, Appendix A). In the Sun's LSR the event took place close to the current position of Tyc~3179-00756-1 (which has a very short 
solid cyan trajectory in Fig.~\ref{chartimage}) but if we subtract the cluster motion we find that the event took place close to its center (non-filled cyan star in the bottom panels of 
Fig.~\ref{chartimage}). The coordinates of the event are known with high precision (the semi-major axis of the two-sigma uncertainty ellipses in the Fig.~\ref{chartimage} is just 
2\arcmin\ long) because they are the result of the intersection of five trajectories.

Most of the plane-of-the-sky momentum of the event in the frame of reference of the cluster was carried away by the Bajamar star and by HD~\num{195965} but with a great imbalance in 
velocities produced by the differences in mass. Hence, the Bajamar star is still in the region (indeed, providing most of the ionizing photons responsible for the North America and Pelican 
nebula) while HD~\num{195965} is $\sim 6^{\rm o}$ away from the location of the event. Each of those two systems was ejected towards opposite quadrants in the reference frame of the cluster 
(as expected if their total linear momentum were close to zero) but when considering the four systems, there is an excess of $\sim$200~M$_\odot$km/s towards the E. Such an excess can be
corrected by tweaking the stellar masses. One possibility to do so is to reduce the mass of Bajamar~A,B by 15-20\% (to 85-90~M$_\odot$) and to increase the mass of Tyc~3157-00918-1 by 
50-80\% (to 5.4-6.5~M$_\odot$) with respect to the values derived in Appendix~A. The first change is within the expected uncertainties of isochrones (especially considering variations in 
age) and yields a lower mass for Bajamar~A that is more consistent with its spectral type. The second change would require Tyc~3157-00918-1 to be a similar-mass binary or a 
pre-main sequence (PMS) star, which is a prediction that can be tested with high-resolution spectroscopy.

In this paper we only have precise tangential velocities $v_{\rm t}$, as 3-D velocities include a radial component that our modelling can only poorly constraint, as we do not have measured
$v_{\rm r}$ and our {\it Gaia}~EDR3 parallaxes provide only limited information about the distance differences. Considering only $v_{\rm t}$, HD~\num{195965} and Tyc~3157-00918-1 are
above the 30~km/s threshold and are therefore runaway systems. The other three systems are well below the threshold and would be only walkaways unless their $v_{\rm r}$ are significantly 
larger than their $v_{\rm t}$. The {\it Gaia}~EDR3 distances do not appear to allow that with the only possible exception of the Bajamar star itself (which has a relatively large distance 
uncertainty) but a large $v_{\rm r}$ in that case would be problematic in terms of balancing momentum in the radial direction.

When we started analyzing these data, our first guess was that the Bajamar and Toronto stars were simultaneously ejected but all of our attempts with {\it Gaia}~EDR3 proper motions were 
unable to place the two systems at the same place at the same time: at their closest approach Toronto was always towards the SW from Bajamar. Nevertheless, we point that the Toronto
star is the brightest {\it Gaia}~EDR3 source in the sample and that, as such, the multiplicative constant applied to its astrometric uncertainties is large 
{\citep{Maiz21}.} 
If it were even larger, it would be possible to place it at the location of the event. Indeed, if we use the TGAS proper motions \citep{Michetal15},
a common event becomes a possibility but we will likely not have a final answer until future {\it Gaia} data releases reduce the uncertainties for systems as bright as Toronto.

\subsection{The Toronto event}

$\,\!$\indent The Toronto event involved (at least) two stellar systems and took place 1.496$\pm$0.044~Ma (with Gaussian weights) or 1.503$\pm$0.051~Ma ago (without Gaussian weights),
with a total possible flight-time range in the 100 Monte Carlo simulations of 1.402-1.643~Ma. Therefore, it likely happened after the Bajamar event but with {\it Gaia}~EDR3 we 
cannot exclude that both events took place simultaneously $\sim$1.6~Ma~ago. The two systems involved are the Toronto star and HDE~\num{227090}, which result in three
stars (Toronto is a spectroscopic binary, \citealt{Willetal01}) of which one or two are massive stars (Toronto~B is a borderline case). In the frame of reference of the cluster, the 
event took place within its boundaries but the uncertainty ellipse in the lower left panel of Fig.~\ref{chartimage} is considerably larger than that of the Bajamar event, as here there are
only two trajectories involved. Most likely, the Toronto event took place to the SW of the Bajamar event but we cannot exclude that they happened at the same location. However, if that
was the case, then the two events could not have taken place simultaneously given the correlations between \tmin, \alphamin, and \deltamin.

\begin{figure}
\centerline{\includegraphics[width=\linewidth]{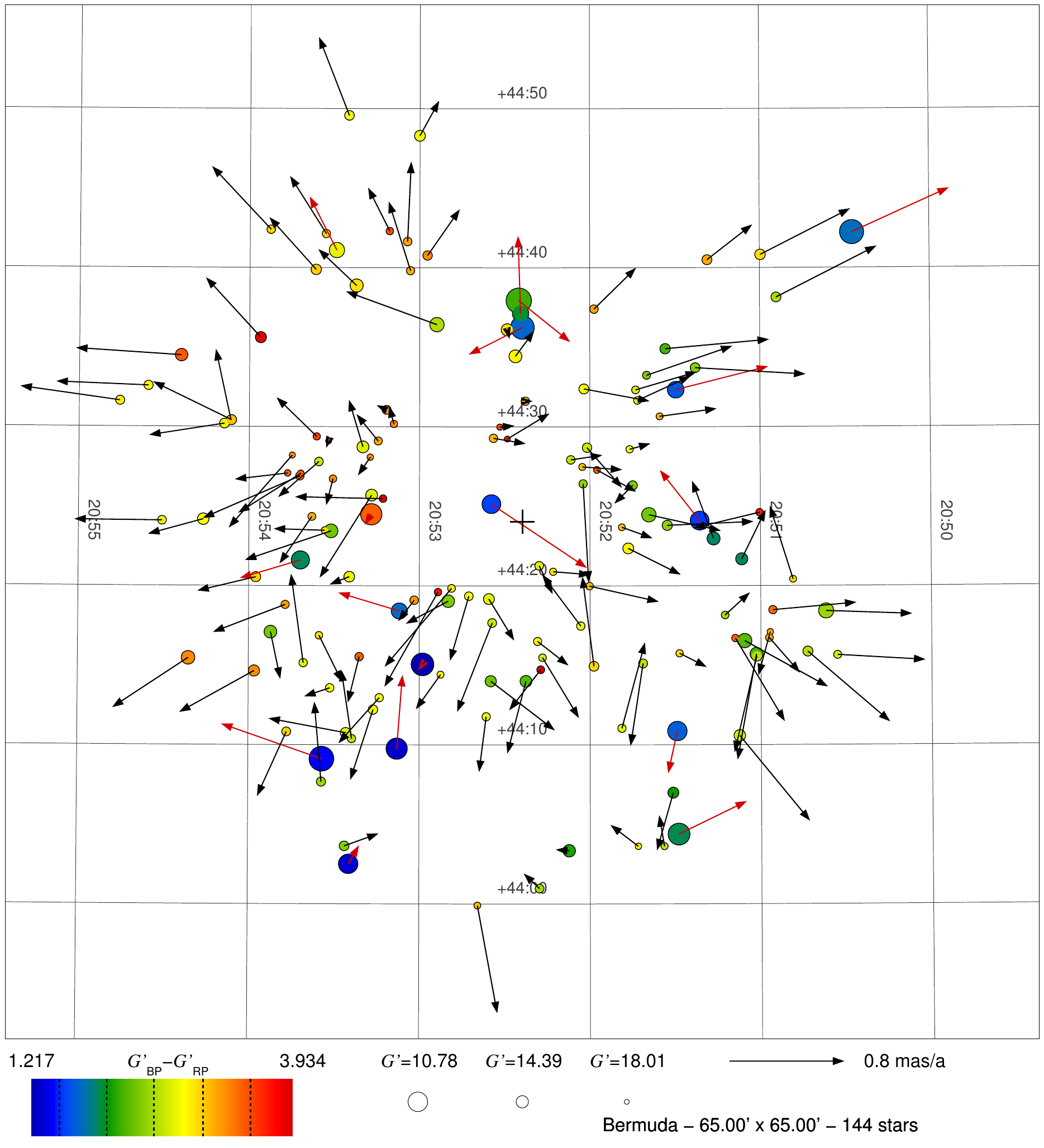}}
 \caption{{\it Gaia}~EDR3 chart of the 144 members of the Bermuda cluster selected in Villafranca~II. Symbol sizes encode \GGc\ (corrected {\it Gaia}~EDR3 magnitude), symbol colors encode 
          \GBPmGRPc\ (corrected {\it Gaia}~EDR3 color), and arrows encode relative proper motions. Red arrows are used for the proper motions of the 17 brightest systems in \GGc. A cross
          indicates the used cluster center.}
\label{pmchart}
\end{figure}

The two systems involved in the Toronto event were ejected in directions nearly perpendicular to the main direction of the previous event as defined by the Bajamar star and HD~\num{195965}.
The Toronto event is even more asymmetric, as HDE~\num{227090} is an intermediate-mass star ejected in the direction nearly opposite to that of Toronto (Fig.~\ref{chartimage}). Such an 
alignment allows for an excellent momentum conservation in the plane of sky, as evidenced in Table~\ref{indivresults}: in both right ascension and declination the total momentum including
the contribution from Toronto~B is within one sigma of zero. The {\it Gaia}~EDR3 distance uncertainties are large (Toronto is a very bright system and HDE~\num{227090} has a bad RUWE) but 
the more massive system appears to have been ejected outwards with respect to the Sun and the less massive system inwards, so (within the current large error bars) momentum can be balanced 
also in the radial direction. A corollary of this result is that the mass ratio between Toronto~A,B and HDE~\num{227090} of 9.1 derived in Appendix~A from the evolutionary masses and the
mass ratio for Toronto~A,B of \citet{Willetal01} agrees very well with the trajectories of the two systems assuming the absence of an additional body.

Considering its value of $v_{\rm t}$, HDE~\num{227090} is a clear runaway. On the other hand, Toronto is just a walkaway in the plane of the sky. Its large uncertainty in distance does 
not allow us to predict its $v_{\rm r}$ but the balance of momentum in the radial direction cannot make it too large. Indeed, \cite{Willetal01} give a heliocentric center of mass radial 
velocity of $-6.4\pm 0.4$~km/s for Toronto and in the Sun's LSR that corresponds to $+5.7$~km/s. Therefore, its velocity with respect to its LSR is in the vicinity of 10~km/s, making the 
Toronto star a walkaway.

\subsection{The HD~\num[detect-all]{201795} event}

$\,\!$\indent The third event took place 1.905$\pm$0.037~Ma ago (with Gaussian weights) or 1.923$\pm$0.062~Ma ago (without Gaussian weights), with a total possible flight-time range in 
the 100 Monte Carlo simulations of 1.787-2.060~Ma. We have been able to identify just two systems that participated in it, HD~\num{201795} and HD~\num{200776}, both of them massive. As it
happened with the previous event, uncertainties in \tmin\ are larger because there are only two systems involved. The information on HD~\num{200776} is relatively limited but 
\citet{Dervetal11} indicate it is an eclipsing binary, something we confirm in Appendix~D, raising the number of ejected stars to three. In the frame of reference of the cluster, the 
HD~\num{201795} event happened close to the location of the previous two but the uncertainty ellipses are too large to determine an accurate separation.

One important difference with the two previous events is that linear momentum cannot be balanced in the plane of the sky, as the two systems were ejected in nearly perpendicular directions. 
If both systems were indeed simultaneously ejected from a 3+ body interaction, one or more stars must be missing. We searched for a third object but could not find any. Could such a third 
object have already exploded as a supernova? For that to happen, the interaction that led to the ejection should have happened when the third object was at least 1-2~Ma old, as otherwise 
there would have been no time for its evolution to lead to a SN explosion. One object that fits the separation, position angle, and approximate mass needed to balance the momentum of the
event is Deneb ($\alpha$~Cyg), which is too bright to be included in {\it Gaia}~EDR3. However, its Hipparcos distance \citep{Maizetal08a} puts it too close to the Sun and, more 
importantly, its relative proper motion \citep{vanL07a} does not point away from the Bermuda cluster.

The value of $v_{\rm t}$ for HD~\num{201795} is already above the 30~km/s threshold to be considered a runway system, independently of its $v_{\rm r}$. HD~\num{200776} is below the threshold
and its $v_{\rm r}$ is small \citep{Dervetal11} and insufficient to push the system over it, so it appears to be a walkway system. The {\it Gaia}~EDR3 distance to HD~\num{200776} puts it beyond the 
North America nebula but with a large uncertainty, so there is no inconsistency with the radial velocity.

\begin{figure*}
\centerline{\includegraphics[width=0.49\linewidth]{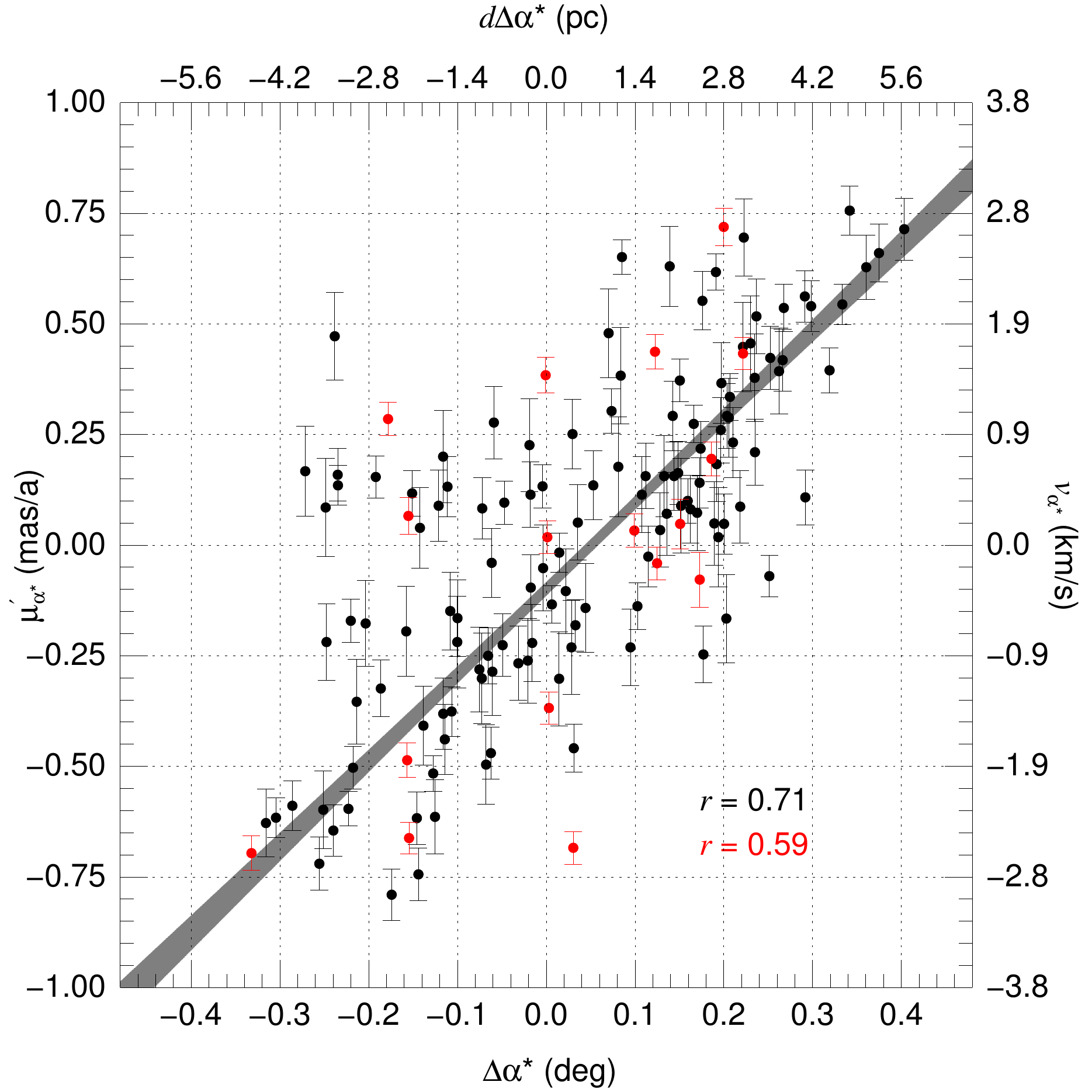} 
            \includegraphics[width=0.49\linewidth]{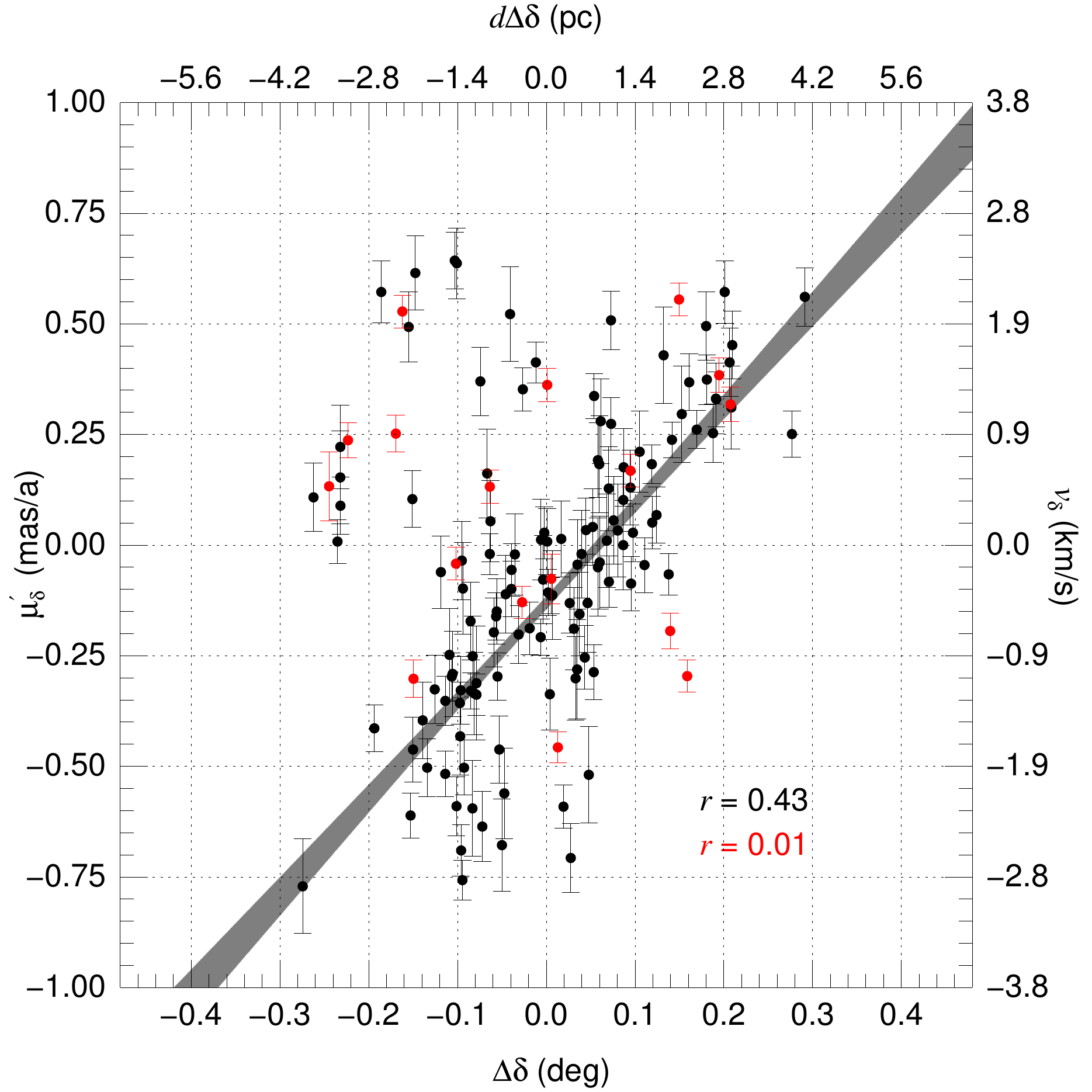}}
\caption{Relative proper motions in right ascension (left) and declination (right) for the 144 members of the Bermuda cluster selected in Villafranca~II as a function of the difference in 
         each coordinate with respect to the cluster center. Red symbols are used for the 17 brightest systems in \GGc\ and black symbols for the 127 fainter systems. The corresponding
         separation and velocity scales are given in the top and right axes, respectively. The correlation factors for the bright and faint systems are also given. The region in gray 
         shows the possible linear fits to the faint systems in each case after rejecting outliers to leave 104 objects (left) and 107 (right).}
\label{pmposition}
\end{figure*}

\subsection{An expanding cluster}

$\,\!$\indent \citet{Kuhnetal20} used {\it Gaia}~DR2 data to detect that the Bermuda cluster (their group D) is expanding with no apparent rotation, as evidenced by their plots of
tangential velocity components as a function of each coordinate (Fig.~17 in that paper) and of proper motion vectors as a function of position in the sky (Fig.~16 in that paper). They also 
noticed that the Bajamar and Toronto stars followed the same velocity-position trends, so ``whatever phenomenon accelerated these stars to their current velocities is also responsible for 
the expansion of the group as a whole'', and provided flight times of $\sim$1.5~Ma for Bajamar and $\sim$1.8~Ma for Toronto. Those values are slightly different from ours, which are obtained
with the more precise (thanks to the longer time baseline of EDR3 with respect to DR2) and accurate (thanks to the astrometric calibrations of 
{\citealt{Maiz21})} 
data available. In this subsection we use the new {\it Gaia}~EDR3 data to reanalyze the \citet{Kuhnetal20} results and in the analysis section below we discuss them in the context of 
how cluster orphanization takes place.

Figure~\ref{pmchart} shoes the {\it Gaia}~EDR3 relative proper motions as a function of position in the sky and Fig.~\ref{pmposition} the behavior of \pmrac\ and \pmdecc\ as a function of
their coordinates. Those figures confirm the finding of \citet{Kuhnetal20} about the expansion of the Bermuda cluster and add new information. In first place, the error bars include the
effect of the covariance (significantly lower than for DR2, \citealt{Maizetal21c}) and the new astrometric calibration of 
{\citet{Maiz21}} 
to demonstrate that the expansion is real and not a measurement effect. 

In second place, there is a significant behavior as a function of \GGc\ magnitude which, though affected by extinction, is highly correlated with
mass (see Appendix~A and Fig.~\ref{CMD}). More specifically, we have divided the 144 Bermuda members selected in Villafranca~II in two groups divided in magnitude: the 17 brightest systems in
one and the 127 faint ones in the other. When we do that, the second group is clearly more correlated in the left plot of Fig.~\ref{pmposition} (expansion in right ascension) and the effect
is even more pronounced in the right plot (expansion in declination), where the second group is still significantly correlated while the bright systems show essentially no correlation 
($r = 0.01$). In the declination plot the effect may be described as the presence of two faint populations, a dominant one that follows the \pmdecc\ vs. $\Delta\delta$ trend and a secondary 
one in the upper left quadrant. When looking at the bright systems, no trend is visible at all and the 17 points show an uncorrelated scatter about the center of the plot. A similar
effect is seen in Fig.~\ref{pmchart}, where the majority of the black arrows point away from the center and their size increase with their distance to it while red arrows do not show a 
clear pattern and instead appear to point in random directions. We have applied a weighted linear regression with an outlier-detection algorithm to fit the data for the faint systems in 
both panels of Fig.~\ref{pmposition} to obtain the expansion time scale in each case. In right ascension we obtain 1.852$\pm$0.077~Ma (104 systems) and in declination we obtain 
1.627$\pm$0.095~Ma (107 systems). Those values are discussed in the analysis section below.

In third place, the Bermuda cluster appears to have been hollowed out in Fig.~\ref{CMD}. There is only one system close to its center and is a bright one, i.e a member of a group that as we 
have seen appears to follow a different pattern with respect to the (majority of) faint systems. This pattern is the opposite of what would be expected for a centrally concentrated cluster 
or even for one with a spherically symmetric density distribution with a flat value near its center. Therefore, it is another piece of evidence in favor of the expansion of the cluster.

\citet{Kuhnetal20} proposed three possible mechanisms for the expansion of the Bermuda cluster: tidal fields, molecular gas dispersion, and initial velocity gradient in their natal cloud. In
the analysis section below we propose a fourth possibility, namely the mass loss associated with the three ejection events in this paper.

\section{Analysis}

$\,\!$\indent The discovery of the multiple stellar ejections from the Bajamar cluster leads to several questions, which are analyzed here. We start with the confidence on our results, we
follow with the effect that the ejections have on the mass function, and we then discuss how cluster orphanization happens. Finally, we discuss other issues related to the discovery and
with possible ways to extend this work.

\subsection{On the reality of the ejection events}

$\,\!$\indent Is it possible what we are seeing is a coincidence and that the ejection events did not really happen but the trajectories just happen to overlap? To answer this question we 
start with the Bajamar event. The first obvious point that needs to be made is that it is not simply that the trajectories overlap but that they coincide at the same position of the sky at 
the same time. This is relevant because the majority of the systems at the range of distances we are sampling have small relative velocities with respect to their LSR that are not sufficient 
to cross several tens of pc in less than 2~Ma, as e.g. HD~\num{195965} and Tyc~3157-00918-1 have done. With that in mind, the probability of back-tracing five trajectories within the sample 
volume (a truncated cone with a radius of $\sim$200~pc and a depth of $\sim$100~pc) to the same location within $\sim$200~AU or less is extremely low. Furthermore, all five systems
are of early-type (O, B, or early-A), including the earliest O-type object in the region (or indeed within 1~kpc of the Sun altogether), another possible O~star, and an early-B star. At a
distance of $\sim$800~pc, the sample is dominated by F and early-G stars with \GG\ fainter than 12 but we see none of those participating. 
Indeed, despite sampling down to $\GG = 14$, the faintest star 
detected in the Bajamar event, Tyc~3575-01514-1, has $\GG = 11.0$, so the ejected systems are not randomly drawn from the sample at all. Additionally, the fact that only minor adjustments to
the evolutionary masses are needed to balance momentum is hard to explain without the alleged simultaneous multiple ejections being a real physical event.  Finally, despite allowing for a 
relative large search area in the sky for the encounter, the location of the event is a region of less than 1~pc in size very close to the cluster center (once its motion has been 
subtracted). Therefore, while it is true that our algorithm is searching essentially in 2-D space (as we do not have much information about depth), the above circumstances make it extremely 
unlikely that the {\it Gaia}~EDR3 data allowing all five systems within such a small volume at the same point in time is just a coincidence.

For the Toronto event we can make some of the same arguments as for the Bajamar event but in this case only two systems are involved and their astrometry has larger uncertainties. As a 
result, the location of the event, though still very close to the cluster center, has a larger uncertainty. Nevertheless, in this case the momentum balance is even more impressive than in 
the previous one. What we have found are two systems ejected in opposite directions (within the uncertainties) and with very different masses ($q = 9.1$) but with velocities that compensate 
for the mass ratio perfectly (within one sigma) to conserve momentum, a feat difficult to accomplish with just two systems. Furthermore, the age of the event is very similar to that of the 
Bajamar event, a connection that points towards a large fraction of the cluster forming simultaneously if ejections take place preferentially within a short time after its birth. Therefore, 
the case for the reality of the Toronto event is still strong. The one point that could change in the future with better astrometry would be the relationship between the Bajamar and Toronto 
events, that is, their degree of spatial and temporal simultaneity.

The HD~\num{201795} event is the one for which the certainty is lower. As for the Toronto event, there are only two systems and that leads to a larger uncertainty on its location (though the
data are still compatible with the event taking place near the cluster center). More importantly, momentum is clearly imbalanced so either a third body is missing or the event did not really
happen. However, we should also point out that this is the only event for which all (two in this case) the detected systems are bright and previously well identified as being of early type
e.g. both were present in the ALS catalog and both are seven-magnitude objects in \GG. As a reference, they are among the 400 brightest (in \GG) systems within 15~degrees of the Bermuda
cluster with measured {\it Gaia}~EDR3 parallaxes between 0.8~mas and 2.0~mas. The importance of the HD~\num{201795} event is that, if real, it extends the time span of the ejection events
$\sim$300~ka in the past from the Bajamar event which, if the ejections really take place a short time after the massive stars are formed, implies that there are several star-forming events
separated by several hundreds of ka. We return to that point below but here we finish by stating that the confirmation of the reality of the HD~\num{201795} event is the most pressing
issue that still needs to be resolved regarding the orphanization of the Bermuda cluster.

\subsection{The IMF and the PDMF}

$\,\!$\indent The preferential ejection of massive stars necessarily modifies the mass function of the cluster. We cannot carry a detailed analysis of the mass function because the presence of
strong differential extinction in the Bermuda cluster (Appendix~A) would require the use of IR information to characterize it and to estimate the completeness of the sample for low-mass stars. 
For that reason, we will center our analysis on the intermediate- and high-mass population.

We use the data in Appendix~A to determine the number of stars in the 12-100~M$_\odot$ and 4-12~M$_\odot$ ranges before and after the ejection events. Of the nine ejected stars listed in 
Table~\ref{chorizos2}, six are in the first range and just one on the second. Adding Bajamar~B and Toronto~B changes those numbers to seven and two, respectively. Of the current eight cluster
members, two are in the 12-100~M$_\odot$ range and six are in the 4-12~M$_\odot$. The number of stars in the first range is very likely to be complete. As for the second range, converting
Tyc~3157-00918-1 into a binary does not change the accounting and the eclipsing companion to HD~\num{200776} is unlikely to be as massive as 4~M$_\odot$. Regarding unaccounted current cluster
members, a look at Fig.~\ref{CMD} indicates the existence of several stars in the 2-4~M$_\odot$ range and it is possible that one or two of them may actually be over 4~M$_\odot$. In summary, 
before the ejections the ratio between the two ranges was 9:8 (or maybe 9:9 or 9:10 if there are missing stars) and the current value after the ejections is 2:6 (possibly 2:7 or 2:8). The total 
ejected mass from Table~\ref{chorizos2} plus Bajamar~B, Toronto~B, and HD~\num{200776} is $\sim$220~M$_\odot$. Considering the possible overestimation of the Bajamar mass discussed above, the 
real value may be slightly smaller but we can establish a lower limit of 200~M$_\odot$.

How does this compare with a Kroupa IMF, which is a multipart power law wth slopes of $-0.3$ ($M < 0.08$~M$_\odot$), $-1.3$ ($0.08$~M$_\odot < M < 0.50$~M$_\odot$), and $-2.3$ 
($0.50$~M$_\odot < M$) \citep{Krou01}? Under that assumption, if there are eight stars in the 4-12~M$_\odot$, one expects 2.4 stars in the 12-100~M$_\odot$ range (changing the number to 
ten stars in the first range raises the second number to 3.0). The observed number of 9 (before ejections) is over three sigmas away from the expected value. Therefore, we conclude that 
{\bf the IMF of the Bermuda cluster was significantly top heavy}, with a measured value of the intermediate/massive-star slope of $\sim -1.3$ instead of the Kroupa value of $-2.3$. 
On the other hand, in the current configuration of the cluster there are six stars in the 4-12~M$_\odot$ range and in that
case a Kroupa mass function predicts 1.8 stars with masses of 12-100~M$_\odot$, which is in good agreement with the observed value of two. This leads to the interesting conclusion that 
{\bf cluster orphanization can transform a top-heavy IMF into a Kroupa-like PDMF.} This discovery is even more relevant when we consider that the Bermuda cluster is the only known case
of a stellar group where stars as massive as Bajamar~A have been recently formed within 1~kpc of the Sun and, therefore, is one of the clusters which we can potentially study better: what we see
here may be hiding somewhere else. 

Another consequence of the assumption of a Kroupa IMF is that if there are eight stars in the 4-12~M$_\odot$, one expects $\sim$655 cluster members in the 0.08-100~M$_\odot$ range. \citet{Kuhnetal20} 
identified 235 members of Bermuda cluster but not reaching down to brown dwarfs so the numbers are consistent within a factor of two. Such a cluster would have a total initial mass of 
$M_{\rm clus} \sim$376~M$_\odot$. The real Bermuda cluster may have had a slightly different initial mass, higher by $\sim$150~M$_\odot$ for being top-heavy and possibly lower if the \citet{Kuhnetal20}
measurement is more complete than expected and the cluster was also initially top heavy when comparing intermediate-mass stars and low-mass ones. A conservative assumption is that the Bermuda
cluster formed 400-500~M$_\odot$ in stars, making it a moderately low-mass cluster. If such a cluster loses $\sim$200~M$_\odot$ soon after its formation, when an additional mass loss in the form of gas 
is also likely to happen, it could be sufficient to unbind it. We explore that possibility in the next subsection.

In Villafranca~II we initially proposed that the Bajamar~star had formed in near isolation. After its characterization as a walkaway star by \citet{Kuhnetal20} we clarified that by near
isolation we included the possibility of the star being born in a low-mass cluster, something that was considered highly unlikely in the \citet{WeidKrou06} scenario but that the results in this
paper confirm. The hidden variable turned out to be the multiple ejections of (preferentially) massive stars significantly changing the PDMF with respect to the IMF.


\subsection{How does cluster orphanization happen?}

$\,\!$\indent The results presented in this paper make the Bermuda cluster a Rosetta stone for understanding massive-star formation and cluster evolution for two reasons: there is no other
stellar cluster that has recently formed stars as massive as Bajamar A within 1~kpc of the Sun and it is extremely young, with episodes younger than 2~Ma. In the previous subsection we
analyzed the consequences that the ejection of the most massive stars of a young cluster has on the PDMF; here we discuss what the process tells us about the formation of massive stars
and the evolution of stellar clusters in their initial stages.

Numerical simulations tell us that the critical parameter to produce dynamical ejections is the cluster radius: ejections are more common in compact clusters, as expected from the increased
likelihood of close encounters at high densities \citep{OhKrou16}. Also, for compact clusters with masses lower than 1000~M$_\odot$ simulations indicate that it is possible to eject all 
stars more massive than 17.5~M$_\odot$ and produce an orphan cluster \citep{Ohetal15}. Those papers assume a Kroupa IMF but, as we have seen above, the Bermuda cluster has a top-heavy IMF and
that increases the possibility of ejections given the existence a higher fraction of massive stars. Therefore, the events described in this paper are likely to have happened in an environment
quite different to that of the present Bermuda cluster, when the cluster was highly compact. In this respect, above we derived expansion time scales for the Bermuda cluster in the 1.6-1.9~Ma
range, values that are compatible with the age of the ejection events\footnote{We note that the cluster expansion is more likely to have experienced a significant deceleration than the
walkaway/runaway stars, given that the cluster members have not moved as far away as the direct participants in the ejection events. This could slightly reduce the time elapsed since the
expansion began and also explain differences between the times associated with the expansion in different directions.}. As we have seen above, the ejected mass 
{may be sufficient} 
to produce such an expansion. All of this points towards a common origin for the ejection events and the cluster expansion at a time when the Bermuda cluster was highly compact.

Nevertheless, the above scenario does not explain two issues: the significant age discrepancy between the HD~\num{201795} event and the other two and the fact that the most massive stars still in
the cluster (along with a few low-mass ones) do not partake on the cluster expansion. One possible explanation is that clusters do not form monolithically but instead do so in a
``conveyor-belt'' fashion \citep{Longetal14,KrumMcKe20}, with two molecular filaments flowing in opposite directions and colliding near-continuously during time scales ($> 1$~Ma) longer than those 
required to form massive stars. This allows large numbers of stars to be formed but the time scale poses a problem: once an O star is born, it will ionize its surroundings and hamper the
subsequent formation of O stars. In the context of the Orion Nebula Cluster, \citet{Krouetal18} suggested a way around: if O stars are preferentially ejected as runaways soon after their births,
new generations of star formation can take place, which brings us back to the Bermuda cluster. We have seen that it has been a highly efficient location for massive star ejections, likely
because of a very small radius, and this could have been the cause for at least two generations of massive-star formation: the first one that produced HD~\num{201795} and HD~\num{200776} and a
second one that produced Bajamar, Toronto, and most of the rest of the stars. Furthermore, after ejecting all the stars form the three events, star formation could have kept going and produce
some additional stars. As those would not have been implicated in the events, they would not follow the cluster expansion pattern and instead form the contaminant population in 
Figs~\ref{pmchart}~and~\ref{pmposition}. The separation of the star formation in the cluster into different episodes separated by hundreds of ka also simplifies the energy balance of the
cluster expansion: the major episode would have been the one that involved Bajamar and Toronto but the more recent one would have added stellar mass to the cluster a posteriori, making the
Bermuda cluster even less massive at the time of the ejections. The high rate of stellar ejections would also explain the anomalous top-heavy IMF: as O stars have not been left around to ionize 
the infalling molecular gas, massive-star formation has continued for a longer period of time than if they had remained in the cluster. In summary, the available data is consistent with 
{\bf star formation taking place in the Bermuda cluster in a conveyor belt fashion with massive stars being ejected and in that way suppressing their negative feedback and leading to a top-heavy IMF.}

{As we discussed in Villafranca~I, there are two views on the formation of OB associations: the traditional (or nurture) view of \citet{LadaLada03} that states that all stars are born in 
clusters but the feedback from protostellar outflows, UV radiation, and stellar winds sweeps the leftover gas and unbinds the stars by lowering the absolute value of their potential energy; and the 
alternative (or nature) view of \citet{Elme10,Wardetal20} that states that star formation is a hierarchical process that happens at very different length scales, leading to stars being born in both
bound (clusters) and unbound (associations) stellar systems. The analysis of \citet{KrumMcKe20} suggests that clusters are formed under the special circumstances of the conveyor belt model when
the configuration is such that the molecular gas in two opposing filaments keeps feeding the same location of star formation for a significant amount of time. In} 
{most circumstances this does not
happen and we end up with multiple sites of star formation in a large cloud. Also, the full mass of a single cloud is never assembled at a single point in time. Recent analyses
of the expansion of OB associations have produced mixed results \citep{Wrig20} but, in general, it is not true that all show a consistent isotropic expansion, tipping the balance towards the
hierarchical model of 
star formation. How does the Bermuda cluster fit in this context? Interestingly, this is a system that appears to have been bound at one point and where the main possible source 
of energy injection into the leftover gas (O stars) was lost early in the process, thus facilitating the continuation of the star formation process through the conveyor belt mechanism. However, the 
cluster eventually lost a significant amount of mass in the form of stars (and, at some point, also gas) and has ended up unbound, so its future is an expanding OB association. How common is 
such a process? We know that the Bermuda cluster is not a run-of-the-mill system, as stars as massive as Bajamar A are seldom found in star-forming regions but in this respect, as in most aspects related 
to the top of the IMF, the sample is scarce. Further analyses with {\it Gaia} of similar systems, looking for evidence of similar ejections, are needed.} 

{A final point about the process of cluster orphanization is that, as already mentioned, we have been unable to detect any ejected star below 3~M$_\odot$ despite having data of
sufficient quality to detect systems down to $\sim$1~M$_\odot$ with moderate extinction. A scenario that explains that is based on the conveyor belt model described above. Star formation in the
Bermuda cluster took place during $\sim$1~Ma in a series of events spaced by hundreds of ka. In each event a combination of low-, intermediate-, and (in some of them) high-mass star 
formation was started but before the next event happened, only the most massive systems had time to reach the ZAMS and the rest were still on the PMS. In the time between events, the newly formed stars 
were in unstable dynamical configurations that eventually led to the three ejection events described in this paper. However, any low-mass stars would have been significantly larger in size (especially
if still on the early phase of their Hayashi tracks) at that point and, if involved in such a dynamical interaction, they could have been disrupted by tidal forces. This would explain the top-heavy mass 
distribution of the ejected objects and would indicate that ejections take place very early in the life of the cluster.} 

\subsection{Other issues}

$\,\!$\indent There are several early-type O stars among the previously confirmed runaway/walkaway stars. \citet{GvarGual11} mention BD~$+$43~3654 (O4~If, \citealt{Maizetal16} and 
$\lambda$~Cep (O6.5~I(n)fp, \citealt{Sotaetal11a}) and \citet{Maizetal18b} mention ALS~4962 (an uncertain ON5~Ifp) and ALS~\num{11244} (O4.5~III(n)(fc)p). However, those are later than the three
stars classified as O2~If*/WN5 or O2~If*/WN6 ejected from Westerlund~2 (\VO{004}): THA~35-II-42 (=~WR~21a), SS~215 (=~WR~20aa), and WR~20c (\citealt{RomLetal11}; Villafranca~I). Outside the
Milky Way, the most massive runaway known is VFTS~682 \citep{Renzetal19a}. To that select club of massive runaways/walkaways we add the Bajamar star.

At what stage do ejections happen? In the numerical simulations of \citet{OhKrou16} most ejections take place before an age of 1~Ma but some happen as late as 3~Ma. However, those
simulations are for a monolithic cluster with 3000~M$_\odot$, so they do not necessarily apply here. If the conveyor-belt scenario above is correct, then most ejections may happen even earlier
than 1~Ma, right after the first massive stars are formed and the cluster is in a highly dynamical state caused by the chaotic interactions at its core \citep{Bateetal03}. This would be
consistent with the agreement between the two CHORIZOS results for Bajamar~A in Appendix A, which imply that the age of the system is not too different from that of the 1.78~Ma isochrone. 
On the other hand, if the reason why the HD~\num{201795} event does not have a balanced momentum is because the third body has already exploded as a SN, then the ejection must have taken place 
at least 1~Ma (and possibly more) after the formation of the progenitor. A further analysis of the system may provide answers to the question but, at the very least, our study sets a lower limit
for the age of the first massive stars in the Bermuda cluster (1.9~Ma ago) and for the age of the first O stars (1.6~Ma ago).

Are there any other stars in the region that may have been ejected from the North America nebula? The Cygnus loop or Veil nebula is a well studied SNR at a distance of 735$\pm$25~pc 
\citep{Feseetal18} to the south of the North America nebula just outside the top panels in Fig.~\ref{chartimage}. Could the progenitor have been ejected from the North America nebula? A 
velocity of $\sim$100~km/s would be needed if the age of the event is 2.0~Ma but a lower value would suffice if it were older (thus placing the north of the first massive stars in the region
even further into the past). \citet{Fangetal17b} already modelled the morphology of the Cygnus loop as produced by a runaway star with characteristics similar to those described here but
we must point out that the direction from the nebula to the SNR is the wrong one to balance the momentum of the HD~\num{201795} event, so if there is a connection then it must have originated 
in a separate event. One prediction we can make is that as the distance is shorter than the value for the Bermuda cluster, chances are that if the progenitor was ejected from the North America
nebula, then it was originally moving towards us. In that case, the southern edge should be closer than northern edge.

{As discussed in Villafranca~I~and~II, other clusters may have experienced orphanization. \VO{012}~S (Haffner~18) is probably the object that deserves to be analyzed next, as its most
massive stars left the cluster $\sim$400~ka ago. \VO{023} (Orion nebula cluster), arguably the best-studied star-forming region that has produced O stars, is another complex case in which objects 
were ejected as early as 2.5~Ma ago \citep{BlaaMorg53,Hoogetal00} and as late as in historical times \citep{Balletal20,Maizetal21g}. In that case the question should not be whether the most massive
star ever formed there is still in the cluster (it apparently is, $\theta^1$~Ori~Ca) but whether at some point the most massive stars were ejected and left the cluster temporarily orphaned (apparently
that happened with the event from 2.5~Ma ago). If anything is clear, that is that the formation of massive stars and their relationship to stellar clusters is so complex that each case needs to be 
analyzed in detail in order to understand it.} 

\subsection{Future work}

$\,\!$\indent There are several avenues that can be explored to confirm and improve the results presented here. The most obvious one is the measurement of the radial velocities, which 
requires multiple epochs to discard the existence of spectroscopic binarity (if single) or the measurement of the systemic velocity $\gamma$ (if multiple) and of a consistent methodology to 
avoid the offsets of several km/s that plague the values for OB stars \citep{Trigetal21}. Second in the list are spectral classifications for those objects without them and possibly detailed 
spectroscopic modelling to refine the stellar parameters. Third, for objects such as Bajamar, the derivation of the spectroscopic orbits. Finally, the improved astrometry from future {\it Gaia} 
data releases should place more astringent values on the properties of the analyzed stars.

\begin{acknowledgements}
{We thank the referee, John Bally, for his suggestions to improve the paper.} 
J.~M.~A. thanks M.~A.~Kuhn for a conversation that gave him the idea to search for stars escaping from the North America nebula and M. Cervi\~no for useful comments.
J.~M.~A. and M.~P.~G. acknowledge support from the Spanish Government Ministerio de Ciencia e Innovaci\'on through grant PGC2018-\num{095049}-B-C22. 
R.~H.~B. acknowledges support from ANID FONDECYT Regular Project \num{1211903} and the ESAC visitors program.
M.~W. acknowledges support from the Spanish Government Ministerio de Ciencia e Innovaci\'on (MICI/FEDER, UE) through grant RTI2018-\num{095076}-B-C21, and 
from the Instituto de Ciencias del Cosmos de la Univ. de Barcelona (ICCUB, Unidad de Excelencia ``Mar\'{\i}a de Maeztu'') through grant CEX2019-\num{000918}-M.
This work has made use of data from the European Space Agency (ESA) mission {\it Gaia} (\url{https://www.cosmos.esa.int/gaia}), 
processed by the {\it Gaia} Data Processing and Analysis Consortium (DPAC, \url{https://www.cosmos.esa.int/web/gaia/dpac/consortium}).
Funding for the DPAC has been provided by national institutions, in particular the institutions participating in the {\it Gaia} Multilateral Agreement. 
The {\it Gaia} data is processed with the the computer resources at Mare Nostrum and the technical support provided by BSC-CNS.
This publication makes use of data products from the {\it Wide-field Infrared Survey Explorer} ({\it WISE}), which is a joint project of the University of California, Los Angeles;
and the Jet Propulsion Laboratory/California Institute of Technology, funded by the National Aeronautics and Space Administration (NASA), where the word ``National'' refers 
to the United States of America. This paper includes data collected by the {\it Transiting Exoplanet Survey Satellite} ({\it TESS}) mission. Funding for the {\it TESS} mission is provided 
by the NASA's Science Mission Directorate.  This research has made extensive use of the \href{http://simbad.u-strasbg.fr/simbad/}{SIMBAD} and 
\href{https://vizier.u-strasbg.fr/viz-bin/VizieR}{VizieR} databases, operated at 
\href{https://cds.u-strasbg.fr}{CDS}, Strasbourg, France. 
\end{acknowledgements}


%

\bibliographystyle{aa} 
\bibliography{general} 

%

%
%

\begin{appendix}

\section{CHORIZOS analysis}

\subsection{Data and methods}

$\,\!$\indent To derive the properties of the systems in this paper from their photometry we use 
the Bayesian code CHORIZOS \citep{Maiz04c}.

{\it Photometric data.} We use six optical photometric bands, three from {\it Gaia}~DR2 and three from {\it Gaia}~EDR3,
$G_{\rm BP,DR2}$ + $G_{\rm DR2}$ + $G_{\rm RP,DR2}$ + $G_{\rm BP,EDR3}$ + $G_{\rm EDR3}$ + $G_{\rm RP,EDR3}$, applying 
the corrections and calibrations of Weiler et al. (in preparation). As discussed there, the effective passbands have changed
slightly from DR2 to EDR3 and additional information can be obtained by combining the two data releases rather than using
just one of them. To do that accurately, one needs to apply several corrections, divide the $G_{\rm BP,DR2}$ and 
$G_{\rm RP,DR2}$ into two $G_{\rm DR2}$ magnitude ranges at a value of 10.87, and do a similar division for the DR2 and EDR3
$G$ photometry at a magnitude of 13.0. Note that the new DR2 photometric calibration is similar but slightly different from
that of \citet{MaizWeil18}, as the EDR3 reanalysis has allowed for a better understanding of the DR2 issues. We also use 
the three 2MASS NIR bands $J_{\rm 2M}$ + $H_{\rm 2M}$ + $K_{\rm 2M}$ \citep{Skruetal06} with the \citet{MaizPant18} 
 calibration. To check for the existence of MIR excesses we have also downloaded their {\it WISE} $W1+W2+W3+W4$ photometry 
\citep{Cutretal13} but we have not included it in the fit itself. As listed in Tables~\ref{chorizos1}~and~\ref{chorizos2}, 
for one system we have discarded the $G_{\rm EDR3}$ photometry because it was incompatible with the other five optical
bands and for another three we have eliminated one or two of the 2MASS bands because their IR excesses extended into the NIR.

\begin{figure}[h!]
\centerline{\includegraphics[width=\linewidth]{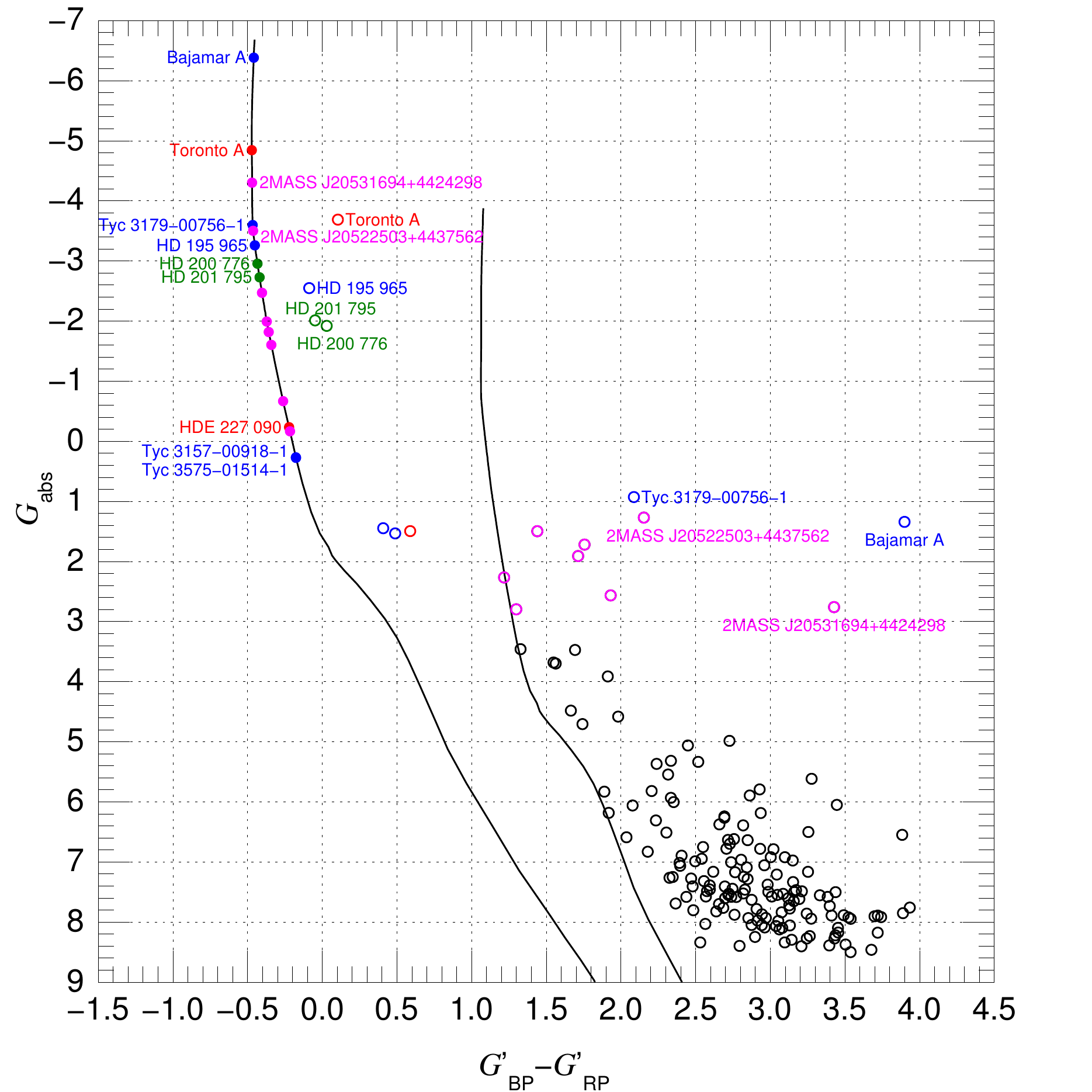}}
\caption{CMD with the extinction-corrected values (filled color circles) from Table~\ref{chorizos2} and extinction-uncorrected ones
         (unfilled color circles). Some systems are labelled to allow for an easier correspondence with Table~\ref{chorizos2}. Different colors are used for the 
         Bajamar (red), Toronto (blue), and HD~\num{201795} (dark green) events and for the eight brightest members of the Bermuda cluster (magenta). 
         The Bajamar~B and Toronto~B components are not plotted but are discussed in the text.
         Unfilled black circles show the extinction-uncorrected measurements for the rest of the Bermuda cluster members selected in Villafranca~II. The two 
         black lines are the 1.78~Ma solar-metallicity Geneva isochrone with no rotation (a) without extinction and (b) with \EBV~=1.0~mag and \RV~=3.0. 
         Magnitudes are from {\it Gaia}~EDR3.}
\label{CMD}
\end{figure}

\begin{table*}
 \caption{Results of the CHORIZOS analysis for the runaway+walkaway sample in this paper using the \Teff-LC grid and 
          $G_{\rm BP,DR2}$ + $G_{\rm DR2}$ + $G_{\rm RP,DR2}$ + $G_{\rm BP,EDR3}$ + $G_{\rm EDR3}$ + $G_{\rm RP,EDR3}$ + 
          $J_{\rm 2M}$ + $H_{\rm 2M}$ + $K_{\rm 2M}$.}
\centerline{
\begin{tabular}{lllllllll}
\hline
Star                      & \mci{\Teff}     & \mci{LC}      & \mci{\EBV}      & \mci{\RV}     & \mci{\chir} & \mci{\AG}       & \mci{\Gabs}        & \mci{$\log L$}  \\
                          & \mci{(kK)}      &               & \mci{(mag)}     &               &             & \mci{(mag)}     & \mci{(mag)}        & \mci{(solar)}   \\
\hline
Bajamar star              &    43.3 (fix.)  & 1.13$\pm$0.04 & 3.600$\pm$0.017 & 2.90$\pm$0.02 & 8.85        & 7.740$\pm$0.018 & $-$6.573$\pm$0.017 & 6.084$\pm$0.007 \\
Tyc~3179-00756-1          &    25.7$\pm$5.7 & 4.74$\pm$0.39 & 1.832$\pm$0.027 & 2.79$\pm$0.06 & 0.68        & 4.325$\pm$0.017 & $-$3.401$\pm$0.017 & 4.334$\pm$0.251 \\
HD~\num{195965}           &    28.1 (fix.)  & 5.08$\pm$0.01 & 0.198$\pm$0.006 & 3.20$\pm$0.17 & 0.16        & 0.644$\pm$0.017 & $-$3.194$\pm$0.016 & 4.250$\pm$0.007 \\
Tyc~3575-01514-1          &    11.7$\pm$1.1 & 5.26$\pm$0.16 & 0.401$\pm$0.024 & 2.88$\pm$0.16 & 1.17        & 1.098$\pm$0.114 & $+$0.432$\pm$0.112 & 1.987$\pm$0.116 \\
Tyc~3157-00918-1          & \phz9.2$\pm$1.7 & 4.93$\pm$0.21 & 0.206$\pm$0.110 & 2.90$\pm$0.29 & 0.10        & 0.570$\pm$0.347 & $+$0.878$\pm$0.346 & 1.689$\pm$0.264 \\
\hline
Toronto star              &    37.9 (fix.)  & 4.51$\pm$0.04 & 0.321$\pm$0.008 & 3.63$\pm$0.16 & 1.06        & 1.166$\pm$0.026 & $-$4.909$\pm$0.025 & 5.240$\pm$0.010 \\
HDE~\num{227090}$^1$      &    12.4$\pm$2.0 & 5.12$\pm$0.24 & 0.453$\pm$0.053 & 3.27$\pm$0.19 & 0.37        & 1.406$\pm$0.227 & $+$0.005$\pm$0.226 & 2.272$\pm$0.160 \\
\hline
HD~\num{201795}           &   27.2 (fix.)   & 5.34$\pm$0.01 & 0.217$\pm$0.006 & 3.17$\pm$0.15 & 0.25        & 0.698$\pm$0.016 & $-$2.712$\pm$0.015 & 4.028$\pm$0.006 \\
HD~\num{200776}           &   24.7 (fix.)   & 5.02$\pm$0.01 & 0.216$\pm$0.012 & 4.23$\pm$0.30 & 1.49        & 0.906$\pm$0.021 & $-$2.837$\pm$0.020 & 3.984$\pm$0.008 \\
\hline
\multicolumn{9}{p{18cm}}{Notes: The first four data columns are the fitted parameters in the grid.
          The fifth column is the reduced $\chi^2$ of the fit. The last three columns are derived quantities and auxiliary 
          parameters. Stars with spectral types in Table~\ref{sample} have their \Teff\ fixed from them. The uncertainties 
          in \Gabs\ and $\log L$ do not include the distance uncertainty i.e. they assume the exact distance taken from 
          Table~\ref{indivresults}. No attempt has been done to separate the effects of the known companions for the Bajamar 
          and Toronto stars. Hence, their LC and $\log L$ values are uncorrected.} \\
\multicolumn{9}{l}{Photometry excluded from the fit. $^1$: $G_{\rm EDR3}$.}
\end{tabular}
}
\label{chorizos1}
\end{table*}

{\it Models.} We use two grids based on the \Teff-LC (luminosity class) grid with solar metallicity from 
\citet{Maiz13a}. One is the full grid in which \Teff\ and LC are the two intrinsic independent parameters and for the other 
one we force the result to be in the 1.78~Ma solar-metallicity Geneva isochrone with no rotation \citep{LejeScha01}, for which
the only intrinsic independent parameter is the stellar initial mass, \mi. The parameter range that is investigated is covered 
by the TLUSTY (hot stars, \citealt{LanzHube03,LanzHube07}) and Munari (cool stars, \citealt{Munaetal05}) parts of the grid, so 
data from both sources are used. In particular, note that the NIR colors from TLUSTY models for late/mid-B stars are 
incorrect by up to several hundredths of a magnitude, so in that region of the spectrum the \citet{Maiz13a} grid uses the 
Munari models for all hot stars.

{\it Distances.} For the systems that are still cluster members we use the distance to the Bermuda cluster from Villafranca~II
(798~pc). For the walkaways/runaways we use the {\it Gaia}~EDR3 individual parallaxes with the calibration of 
{\citet{Maiz21}} 
and use the prior of \citet{Maiz01a,Maiz05c} with the parameters of 
\citet{Maizetal08a}. In particular, for the walkaways we assume they are part of the disk population and for the runaways
that they are part of the halo or runaway population.

{\it Other parameters.} For the full \Teff-LC grid runs we fix the distance and fit the other four parameters (\Teff, LC,
amount of extinction \EBV, and type of extinction\footnote{See \citet{Maiz04c,Maiz13b} and \citet{MaizBarb18} for an 
explanation of why one needs to use monochromatic quantities instead of band-integrated ones to characterize extinction.}
\RV) unless the star has a known spectral type, in which case we also fix \Teff. To establish \Teff\ from the spectral type 
we use a version of the \citet{Martetal05a} calibration extended to later spectral
types and adapted to the subtypes changes of \citet{Sotaetal11a,Sotaetal14,Maizetal16}. For the isochrone grid fits we also 
fix the distance and fit the other three parameters (\mi, \EBV, and \RV). For all runs we use the family of extinction laws of 
\citet{Maizetal14a}. 

{\it Why two grids?} The reasons for using two grids (full \Teff-LC and isochrone) is that each one has its pros and cons and 
that two independent results allow for additional checks. The most accurate results are those for the \Teff-LC grid when 
\Teff\ is fixed from the spectral type. In that case if a system is a hidden multiple with similar spectra types the effect is 
to make the system of a higher luminosity class without altering other parameters, as colors for OB stars have a weak 
dependence on luminosity. On the other hand, if \Teff\ is not fixed because no spectral type is available, the resulting 
uncertainties can be rather large (see e.g. Tyc~3179-00756-1 in Table~\ref{chorizos1}) as the filter combination used 
provides some \Teff\ discrimination but not much (Ma{\'\i}z Apell\'aniz et al. in preparation). The situation becomes 
worse for stars with high extinction and for that reason cluster systems are fitted only using the isochrone grid 
(Table~\ref{chorizos2}) and not the full \Teff-LC grid (Table~\ref{chorizos1}). The isochrone grid fits works very well if
the age is well known and if there are no vertical shifts 
in magnitude caused by distance errors or hidden companions. For that reason,
for the isochrone results for the Bajamar and Toronto stars in Table~\ref{chorizos2} we subtracted the effect of their companions
by adding 0.18~mag and 0.05~mag, respectively, to all magnitudes, with the values estimated from the available spectroscopy and 
the expected differences in spectral types. Using an isochrone of the wrong age can create significant biases in some parameters
and the resulting uncertainties can be underestimated, as it is the case for \mi, \Teff, and $\log L$ in Table~\ref{chorizos2}.

\subsection{Stellar results}

$\,\!$\indent We analyze the CHORIZOS results for the systems in Tables~\ref{chorizos1}~and~\ref{chorizos2} dividing them by the event they
belong to and finishing with the ones that are still cluster members. The extinction characteristics are discussed in the next subsection.

{\it Bajamar event.} The Bajamar star, the most massive system in this paper, shows an excellent consistency in \Teff\ between the two runs, indicating that its age cannot be 
too different from the 1.78~Ma of the isochrone, and a value of 1.13 for the luminosity class, indicating that it has already evolved from the ZAMS. The differences 
in \Gabs\ between the two runs can be adscribed to the second one referring only to the A component (of spectral type O3.5~III(f*)). In conjunction with the mass ratio of 
2.2$\pm$0.5 derived below, these results indicate that the Bajamar star is a binary with masses of $\sim$73~M$_\odot$ and 33$\pm$8~M$_\odot$, the latter within the range expected for
an unevolved O8 dwarf\footnote{The values quoted here are evolutionary masses, as they are derived from isochrones. It is beyond the scope of this paper to compare them with 
Keplerian or spectroscopic masses, as we do not have the data at hand to do so but see the text for a discussion on how the values are compatible with the momentum balance of the
events.}. The first run results for Tyc~3179-00756-1 have relatively large uncertainties but those from the second run point out towards the object being an O9.7/B0 
dwarf if it is a single star of age younger than 2~Ma. This could be a third O star ejected from the Bermuda cluster, something that should be confirmed spectroscopically. The two
runs show similar results for HD~\num{195965}, albeit slightly hotter for the isochrone run, which translates into a higher $\log L$ due to the larger bolometric
correction. The last two systems involved in the event, Tyc~3575-01514-1 and Tyc~3157-00918-1, have results compatible with being late-B or early-A dwarfs. Tyc~3575-01514-1 has a MIR 
excess in the W3 and W4 {\it WISE} bands. The results for the two runs for Tyc~3157-00918-1 are somewhat discrepant, with those of the first run indicating a lower \Teff. One possible way to
reconcile them would be if the system is actually composed of two similar early-A stars, as that configuration will lead to the \Teff\ of the first run but a higher luminosity that would 
be confused in the second run with a higher \Teff. Alternatively, it could be a PMS star of A type. As discussed in the text, in either case the total mass of Tyc~3157-00918-1 would be higher 
than the value in Table~\ref{chorizos2}, which assumes a single MS star.

{\it Toronto event.} The Toronto star shows a good agreement in \Teff\ between the two runs, with a slightly higher value in the second one but within the uncertainties
expected from the \Teff-spectral type calibration \citep{Maizetal14a}, also pointing towards an age close to that of the isochrone. Assuming the mass ratio of 4 from
\citet{Willetal01} leads to a system composed of a $\sim$33~M$_\odot$ and an $\sim$8~M$_\odot$ stars, the second most massive of the ones analyzed in this paper. 
HDE~\num{227090} is an intermediate-mass star with a lower value of \Teff\ using the \Teff-LC grid than using the isochrone grid. Its predicted spectral type is mid/late-B.
Note that the $G_{\rm EDR3}$ magnitude of HDE~\num{227090} is incompatible with the other five {\it Gaia} photometric points and has been excluded from the analysis.

\begin{table*}
 \caption{Results of the CHORIZOS analysis for the systems in this paper using the
          1.78~Ma Geneva isochrone grid with no rotation and $G_{\rm BP,DR2}$ + $G_{\rm DR2}$ + $G_{\rm RP,DR2}$ + $G_{\rm BP,EDR3}$ + 
          $G_{\rm EDR3}$ + $G_{\rm RP,EDR3}$ + $J_{\rm 2M}$ + $H_{\rm 2M}$ + $K_{\rm 2M}$.}
\centerline{
\begin{tabular}{lllllllll}
\hline
Star                          & \mci{\mi}         & \mci{\EBV}      & \mci{\RV}     & \mci{\chir} & \mci{\AG}       & \mci{\Gabs}        & \mci{\Teff}     & \mci{$\log L$}  \\
                              & \mci{(solar)}     & \mci{(mag)}     &               &             & \mci{(mag)}     & \mci{(mag)}        & \mci{(kK)}      & \mci{(solar)}   \\
\hline
Bajamar star A                &    72.96$\pm$0.54 & 3.606$\pm$0.016 & 2.89$\pm$0.02 & 8.91        & 7.733$\pm$0.017 & $-$6.384$\pm$0.016 & 42.75$\pm$0.19  & 5.991$\pm$0.005 \\
Tyc~3179-00756-1              &    18.43$\pm$0.20 & 1.858$\pm$0.012 & 2.86$\pm$0.03 & 0.54        & 4.524$\pm$0.024 & $-$3.599$\pm$0.023 & 33.09$\pm$0.11  & 4.598$\pm$0.012 \\
HD~\num{195965}               &    15.51$\pm$0.12 & 0.212$\pm$0.006 & 3.29$\pm$0.16 & 0.15        & 0.712$\pm$0.017 & $-$3.261$\pm$0.017 & 30.97$\pm$0.08  & 4.380$\pm$0.009 \\
Tyc~3575-01514-1              & \phz3.57$\pm$0.04 & 0.431$\pm$0.006 & 3.06$\pm$0.09 & 0.96        & 1.258$\pm$0.025 & $+$0.275$\pm$0.025 & 13.74$\pm$0.09  & 2.189$\pm$0.019 \\
Tyc~3157-00918-1              & \phz3.58$\pm$0.03 & 0.368$\pm$0.006 & 3.35$\pm$0.10 & 0.94        & 1.187$\pm$0.019 & $+$0.267$\pm$0.019 & 13.77$\pm$0.07  & 2.196$\pm$0.015 \\
\hline
Toronto star A                &    33.10$\pm$0.42 & 0.325$\pm$0.007 & 3.55$\pm$0.01 & 0.85        & 1.164$\pm$0.024 & $-$4.845$\pm$0.023 & 39.36$\pm$0.11  & 5.266$\pm$0.013 \\
HDE~\num{227090}$^1$          & \phz4.46$\pm$0.04 & 0.502$\pm$0.007 & 3.44$\pm$0.08 & 0.28        & 1.648$\pm$0.022 & $-$0.235$\pm$0.021 & 15.81$\pm$0.09  & 2.531$\pm$0.013 \\
\hline
HD~\num{201795}               &    12.40$\pm$0.09 & 0.221$\pm$0.006 & 3.18$\pm$0.15 & 0.22        & 0.716$\pm$0.017 & $-$2.729$\pm$0.016 & 28.10$\pm$0.09  & 4.085$\pm$0.009 \\
HD~\num{200776}               &    13.66$\pm$0.14 & 0.241$\pm$0.011 & 4.26$\pm$0.27 & 1.44        & 1.025$\pm$0.023 & $-$2.956$\pm$0.023 & 29.33$\pm$0.11  & 4.212$\pm$0.013 \\
\hline
2MASS~J20531694$+$4424298$^2$ &    25.12$\pm$0.34 & 3.036$\pm$0.017 & 3.01$\pm$0.02 & 2.64        & 7.056$\pm$0.026 & $-$4.304$\pm$0.025 & 36.72$\pm$0.14  & 4.970$\pm$0.015 \\
2MASS~J20522503$+$4437562     &    17.20$\pm$0.14 & 1.872$\pm$0.011 & 2.99$\pm$0.02 & 2.06        & 4.770$\pm$0.018 & $-$3.503$\pm$0.017 & 32.19$\pm$0.09  & 4.510$\pm$0.010 \\
Tyc~3179-00733-1              &    11.13$\pm$0.11 & 1.449$\pm$0.009 & 3.30$\pm$0.03 & 1.58        & 4.188$\pm$0.023 & $-$2.472$\pm$0.023 & 26.71$\pm$0.10  & 3.931$\pm$0.013 \\
2MASS~J20512895$+$4404230$^3$ & \phz9.10$\pm$0.30 & 1.533$\pm$0.022 & 3.42$\pm$0.09 & 0.49        & 4.554$\pm$0.073 & $-$1.993$\pm$0.072 & 24.14$\pm$0.38  & 3.647$\pm$0.044 \\
Tyc~3179-00615-1              & \phz8.45$\pm$0.08 & 1.228$\pm$0.008 & 3.00$\pm$0.03 & 2.03        & 3.313$\pm$0.021 & $-$1.819$\pm$0.021 & 23.12$\pm$0.11  & 3.539$\pm$0.013 \\
2MASS~J20522374$+$4436151     & \phz7.73$\pm$0.07 & 1.510$\pm$0.009 & 2.68$\pm$0.03 & 1.43        & 3.510$\pm$0.022 & $-$1.605$\pm$0.021 & 22.08$\pm$0.10  & 3.407$\pm$0.013 \\
V2022~Cyg$^2$                 & \phz5.36$\pm$0.10 & 0.960$\pm$0.009 & 3.34$\pm$0.07 & 1.94        & 2.923$\pm$0.039 & $-$0.668$\pm$0.039 & 17.65$\pm$0.17  & 2.820$\pm$0.028 \\
2MASS~J20530782$+$4409456     & \phz4.33$\pm$0.05 & 1.005$\pm$0.008 & 3.27$\pm$0.04 & 0.89        & 2.956$\pm$0.025 & $-$0.167$\pm$0.024 & 15.51$\pm$0.11  & 2.490$\pm$0.016 \\
\hline 
\multicolumn{9}{p{19cm}}{Notes: For the Bajamar star a magnitude difference of 0.18
          has been applied to all filters to account for the presence of the companion (hence the A in the name). Similarly, for the 
          Toronto star a magnitude difference of 0.05 has been used. The last eight entries correspond to the 
          eight brightest (in $G$) members of the Bermuda cluster selected in Villafranca~II. The first three data columns are the 
          fitted parameters in the isochrone grid. The fourth column is the reduced $\chi^2$ of the fit. The last four columns are 
          derived quantities and auxiliary parameters. For the runaway/walkaway systems the distance from Table~\ref{indivresults} is used while
          for the Bermuda cluster members a distance of 798~pc is assumed. All fits assume the systems are single or that any hidden 
          companions do not contribute significantly to the photometry and that there is no uncertainty in the distance or the cluster age, 
          so the real uncertainties should be significantly larger than the formal ones for quantities such as \mi, \Teff, or $\log L$.} \\
\multicolumn{9}{l}{Photometry excluded from the fit. $^1$: $G_{\rm EDR3}$. $^2$: $K_{\rm 2M}$. $^3$: $H_{\rm 2M}$ + $K_{\rm 2M}$.}
\end{tabular} 
}
\label{chorizos2}
\end{table*}

{\it HD~\num[detect-all]{201795} event.} The two systems involved in this event have results in the two runs that are consistent with both being early-B dwarfs. In the case of 
HD~\num{201795} the consistency between the two runs is excellent. Those of HD~\num{200776} are slightly discrepant, with the Table~\ref{chorizos2} results pointing towards higher
\Teff\ values than those expected from its spectral type. The discrepancy could be resolved in two ways: if the star has a spectral type of B0.2-B0.5~V instead of the published one
(in this case we do not currently have GOSSS or \lili\ data) or if the star turns out to be closer to us by 50-100~pc, something that is within the uncertainties of the current 
{\it Gaia}~EDR3 distance and indeed is suggested by the radial velocity measurement of \citet{Dervetal11} assuming that the two systems were close to the current distance to the 
Bermuda cluster at the time of the event. The mass ratio of 5.2 derived by \citet{Dervetal11} indicates that the companion is an intermediate-mass object of $\sim$2.6~M$_\odot$ and,
therefore, has only a minor contribution to the combined light output (Appendix~D and Fig.~\ref{TESS}).

{\it Cluster members.} The brightest cluster members of the Bermuda cluster selected in Villafranca~II have been analyzed using the isochrone grid. They all have significant
extinctions (see below) and have not been studied in much detail due to the recent identification of the cluster. They appear to be two stars with masses of 12-25~M$_\odot$, 
three with masses of 8-12~M$_\odot$, and another three with masses of 4-8~M$_\odot$. Three of them, 2MASS~J20531694$+$4424298, 2MASS~J20512895$+$4404230, and V2022~Cyg have IR
excesses that reach into the NIR and for another two, 2MASS~J20522503$+$4437562 and Tyc~3179-00733-1, the excess is visible but only in the W3 and W4 {\it WISE} bands. The most
interesting object among the cluster members is 2MASS~J20531694$+$4424298, the most extinguished target of the eight and with a photometry consistent with being a late-type O star.
However, such an identification needs a spectroscopic confirmation, especially considering the anomalous SED that this object has in the IR. A second object, 
2MASS~J20522503$+$4437562 is another potential O9.7/B0 star that deserves spectroscopic observations to confirm such spectral type.

\subsection{Extinction results}

$\,\!$\indent The systems analyzed in Tables~\ref{chorizos1}~and~\ref{chorizos2} show a wide range of extinctions (see also Fig.~\ref{CMD}). Those still in the cluster or in its 
neighborhood (the Bajamar star and Tyc~3179-00756-1) have large or very large extinctions while those that have been ejected at large separations have low values of \EBV\ around 
0.2~mag, which are typical for objects at that distance in this region of the Milky Way (see Fig.~8 in \citealt{MaizBarb18}). 

The Bajamar and Toronto stars had been previously analyzed with different data sets in \citet{Maizetal21a} and \citet{MaizBarb18}, respectively. The new results are in reasonable
agreement with the previous ones but uncertainties are lower now thanks to the better quality and improved calibration of the {\it Gaia} photometry. Nevertheless, we should point
out that while the \chir\ values are close to 1.0 for low- and intermediate- extinction objects, the three most extinguished systems have higher values, with the worst case being the
Bajamar star itself, which is the most extinguished of all. As pointed out in \citet{Maizetal21a}, such an increase of \chir\ with extinction is a sign that the family of extinction
laws of \citet{Maizetal14a} is starting to lose validity for objects as extinguished as the Bajamar star (likely as a result of the wrong slope in the NIR, see
\citealt{Maizetal20a}) but we currently do not have a better alternative, as the other popular choices of \citet{Cardetal89} and \cite{Fitz99} fare even worse \citep{MaizBarb18}.

\addtocounter{section}{+1}
\addtocounter{figure}{-1}
\begin{figure*}[h!]
\centerline{\includegraphics[width=\linewidth]{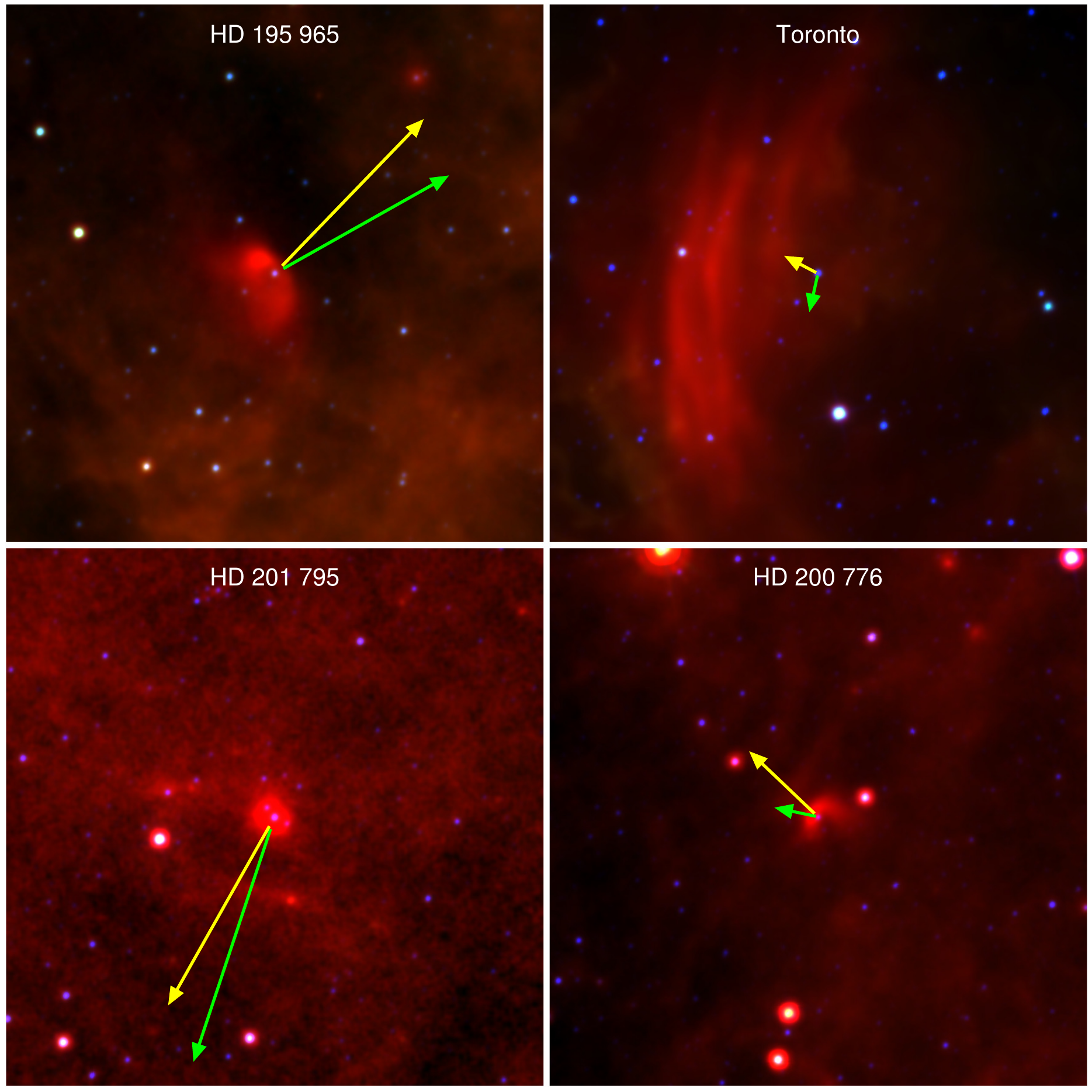}}
 \caption{{\it WISE} W4+W3+W2 RGB mosaics for four walkaway/runaways systems. Each field is $20\arcmin\times20\arcmin$ (4.7~pc~$\times$~4.7~pc at a distance of 798~pc) and is oriented with 
          north toward the top and east toward the left. In each mosaic the runaway candidate is at the center and the arrows show the absolute (green) and the relative-to-the cluster
          proper motions (yellow).}
\label{WISE}
\end{figure*}
\addtocounter{section}{-1}

The results in this paper also confirm other conclusions of \citet{MaizBarb18} and \citet{Maizetal21a}. Extinction can vary greatly within a cluster not only in amount but also in
type of extinction, so using a single value of \RV\ can lead to significant biases. In the case of the North America nebula (excluding runaway systems already far from it) we see
values of \RV\ that range from $\sim$2.7 to $\sim$3.6. Furthermore, the changes are clearly associated with the nebular morphology. Objects behind dense molecular clouds (e.g. 
the Bajamar star or Tyc~3179-00756-1) tend to have large values of \EBV\ and small ones of \RV\ while objects immersed in the \HII\ region (e.g. the Toronto star) tend to have
small values of \EBV\ and large ones of \RV. As we proposed in \citet{MaizBarb18}, extinction in \HII\ regions appears to be produced mostly by large dust grains due to the
selective destruction of small grains, which are instead abundant in molecular clouds.

\section{Bow shocks}

\addtocounter{section}{+1}
\begin{figure*}[h!]
\centerline{\includegraphics[width=0.49\linewidth]{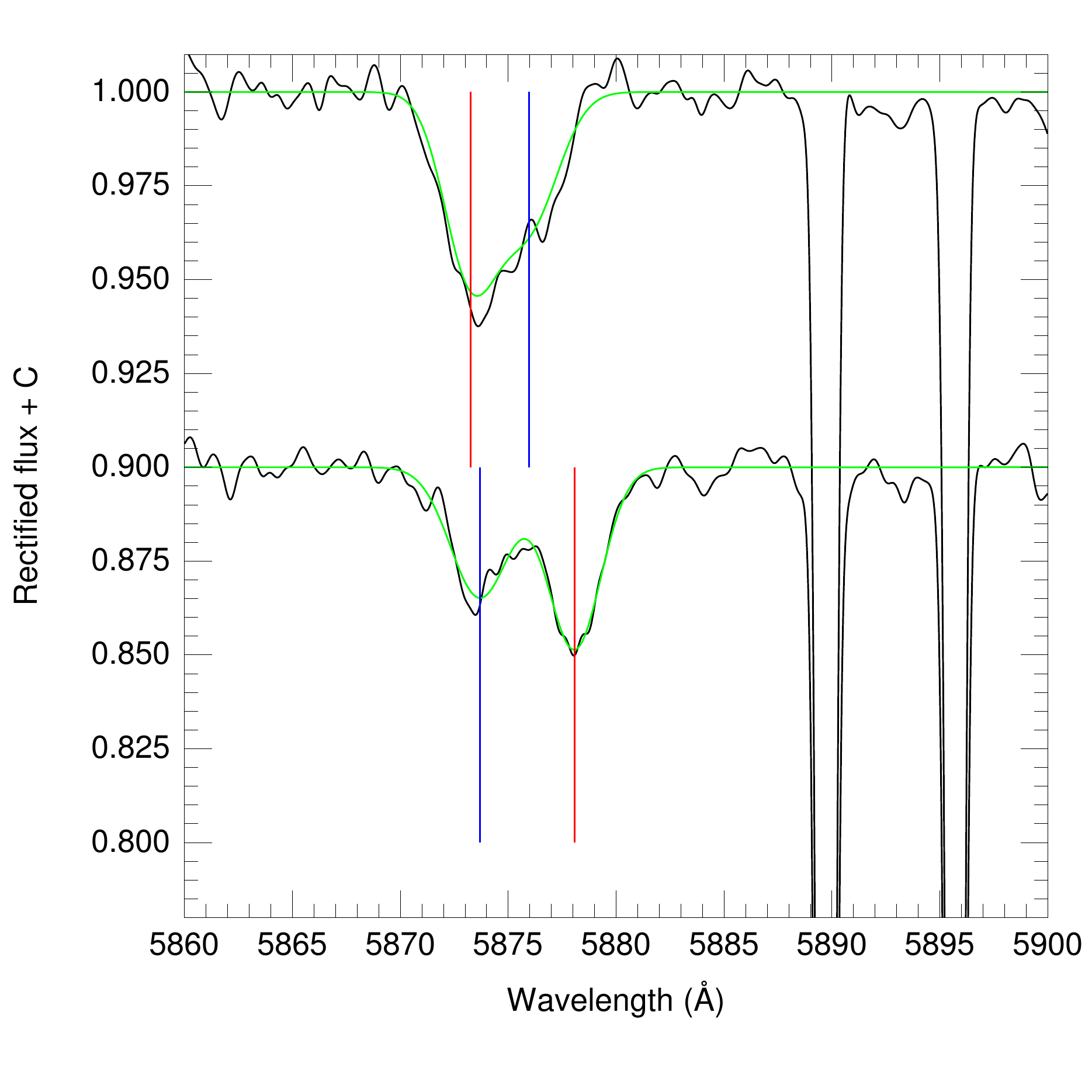} \
            \includegraphics[width=0.49\linewidth]{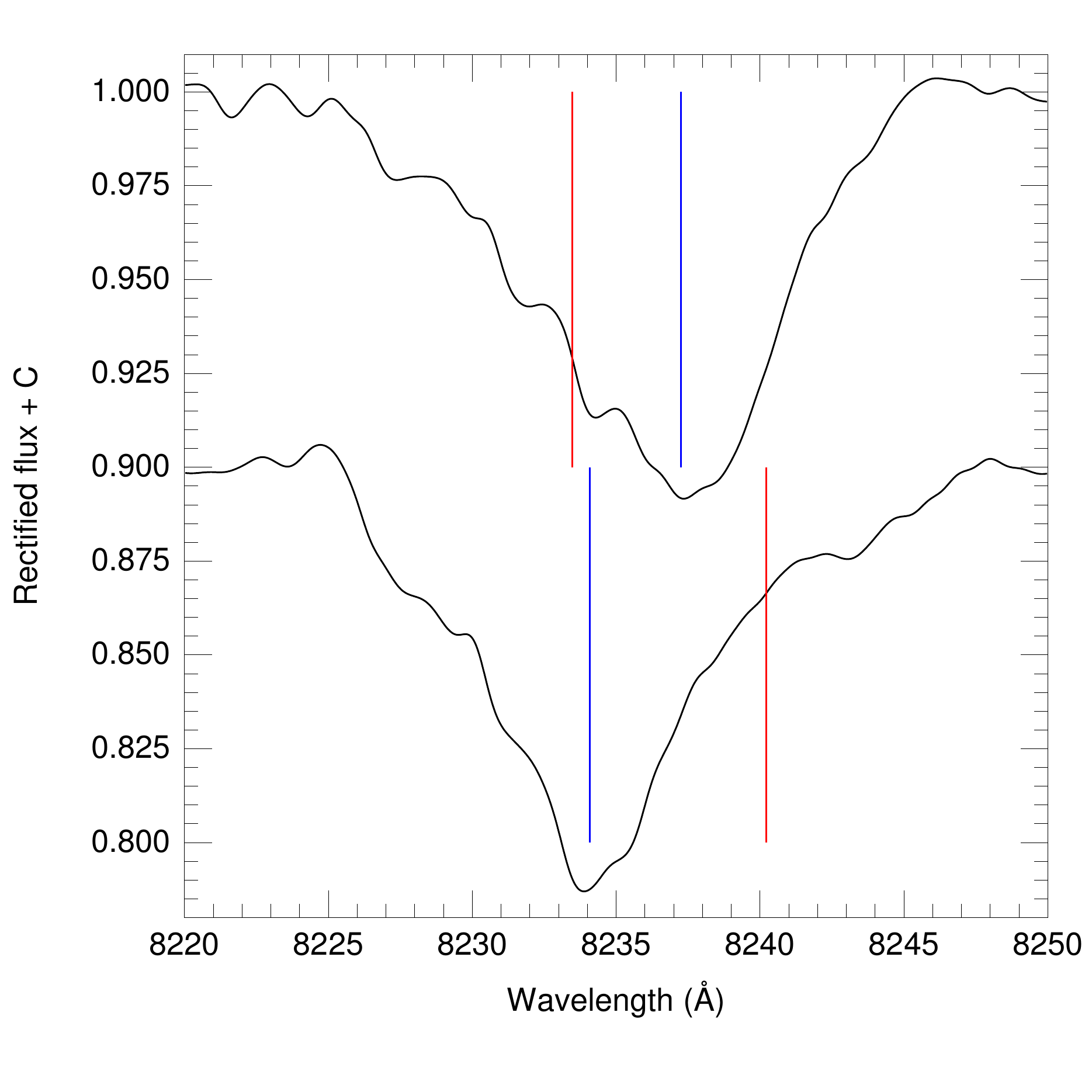}}
 \caption{(left) Two \lili\ epochs of the \HeI{5875.6} and \NaId{5889.951,5895.924} region for the Bajamar star. A double Gaussian profile (green line) is fit to the \HeI{5875.6} 
          absorption profile, with the blue vertical lines marking the A spectroscopic component and red vertical lines the B spectroscopic component. (right) The same two spectra
          in the region of the \HeII{8236.8} line. The blue and red vertical lines mark the expected positions as derived from the velocities of the \HeI{5875.6} fit. In both panels
          the spectral resolution has been degraded to $R$~=~\num{10000} to better visualize the spectra. Telluric lines have been removed following \citet{Gardetal13}.}
\label{Bajamar}
\end{figure*}
\addtocounter{section}{-1}

$\,\!$\indent We have searched for bow shocks created by the walkaway and runaway systems in their {\it WISE} W3+W4 images, 
as such structures can be seen in the {\it WISE} images if the conditions
are appropriate \citep{Maizetal18b}. The Bajamar star and (the relatively nearby in the plane of the sky) Tyc 3179-00756-1 are immersed in a bright MIR nebulosity, 
likely caused by the warm dust on the far side of the Atlantic Ocean molecular cloud heated by them, that does not allow for any faint structure
such as a bow shock to be detected. On the other hand, the other four walkaways/runaways that are known to be of spectral type O or early-B do show bow-shock-like structures 
(Fig.~\ref{WISE}). The O star (Toronto) displays the largest one, followed by the earliest B-type, HD~\num{195965}, as expected from their wind strengths. The bow shocks are better
aligned with the relative proper motions than with the absolute ones (especially so for the Toronto star and HD~\num{200776}), as expected if the ISM in the North America nebula and
its surroundings have similar proper motions to that of the Bermuda cluster. These detections and their orientation confirm the likelihood of the ejections of these four systems from the 
Bermuda cluster.

\section{On the SB2 nature of the Bajamar star}

$\,\!$\indent \citet{Maizetal16} identified the Bajamar star as an SB2 and indicated that some epochs showed velocity differences of $\sim$300~km/s between the He\,{\sc i} and
He\,{\sc ii} lines. No SB2 orbit has been published at this time, most likely because of the difficulty of studying this system. The spectra is dominated by the H and He\,{\sc ii} 
lines but the extremely high extinction makes the blue-violet region difficult to access. The motion of the primary is easy to track with the He\,{\sc ii} lines (right panel of
Fig~\ref{Bajamar}), which are dominated by the hotter (A) component and to which the cooler (B) one contributes 10-15\% of the total equivalent widths and is seen only as 
wings to the profile when the velocity separation is large. To see both components clearly we need to use one of the (weak) He\,{\sc i} lines, where the much higher strength of the B
component is sufficient to overcome the magnitude difference and have a similar (or even larger) equivalent width as the A component (left panel of Fig.~\ref{Bajamar}). However, 
there are few He\,{\sc i} lines available and obtaining good-quality data is time expensive: each of the epochs in Fig.~\ref{Bajamar} required 30-minute integrations on a 3.5~m
telescope.

We are currently obtaining GOSSS and \lili\ observations of the Bajamar star to derive its SB2 orbit but we still require further epochs before we can be certain about the result. Such
an orbit will be included in one of the future installments of the MONOS project \citep{Maizetal19b,Trigetal21}. Nevertheless, we can already present two preliminary results here: the
period is  
{around two} 
weeks and the mass ratio is 2.2$\pm$0.5 (Fig.~\ref{Bajamar}).

\section{TESS light-curve analysis}

\begin{figure*}
\centerline{\includegraphics[width=0.49\linewidth]{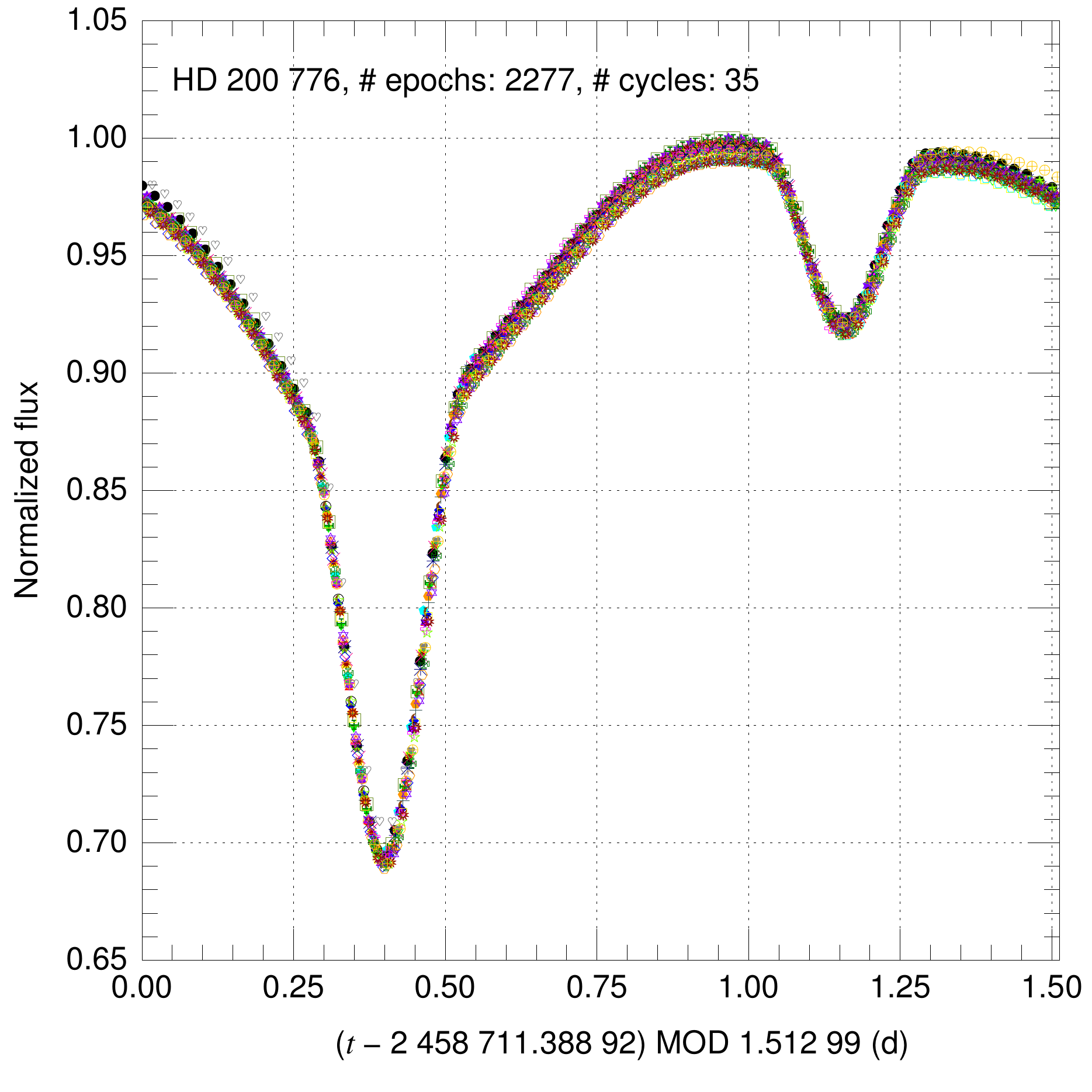} \
            \includegraphics[width=0.49\linewidth]{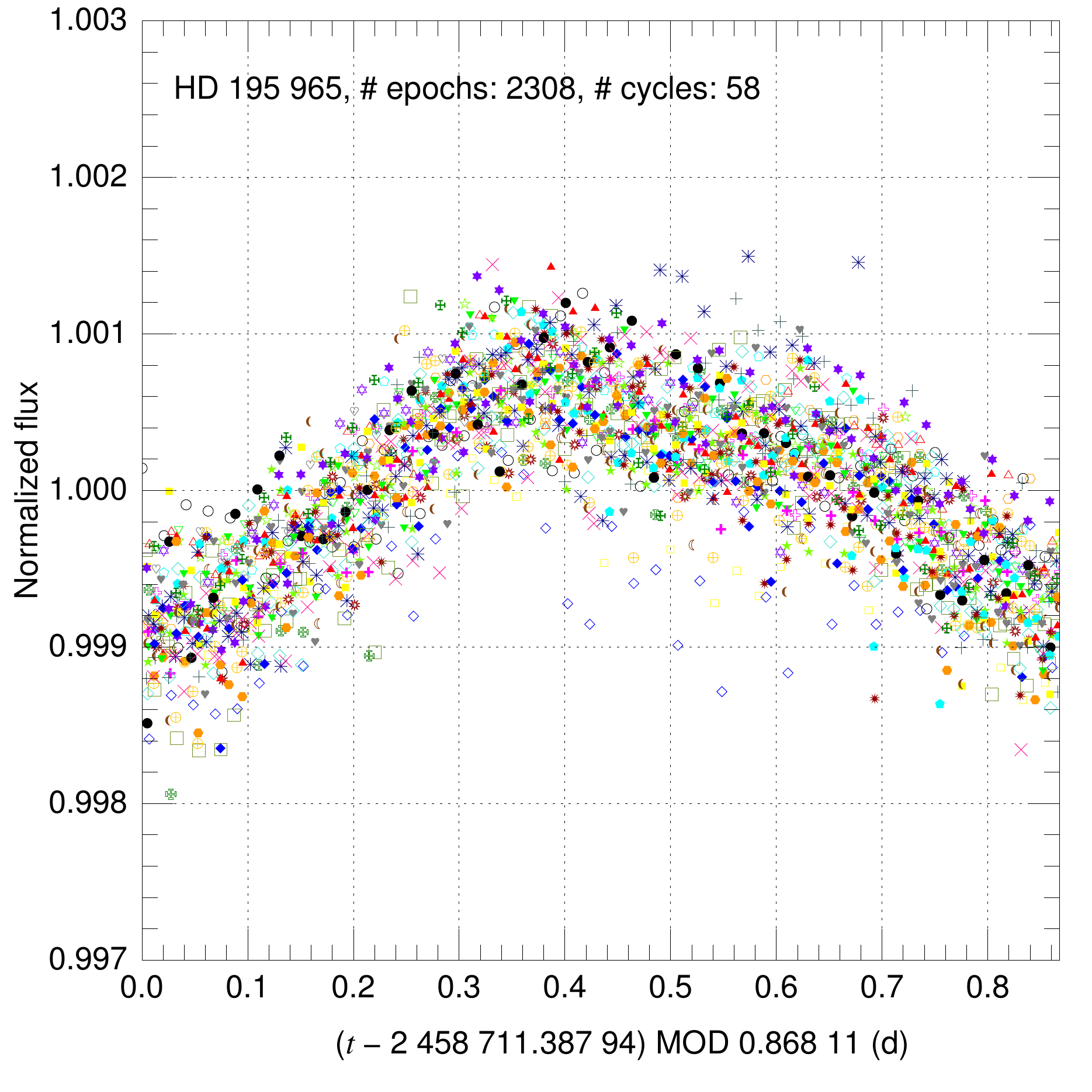}}
\centerline{\includegraphics[width=0.49\linewidth]{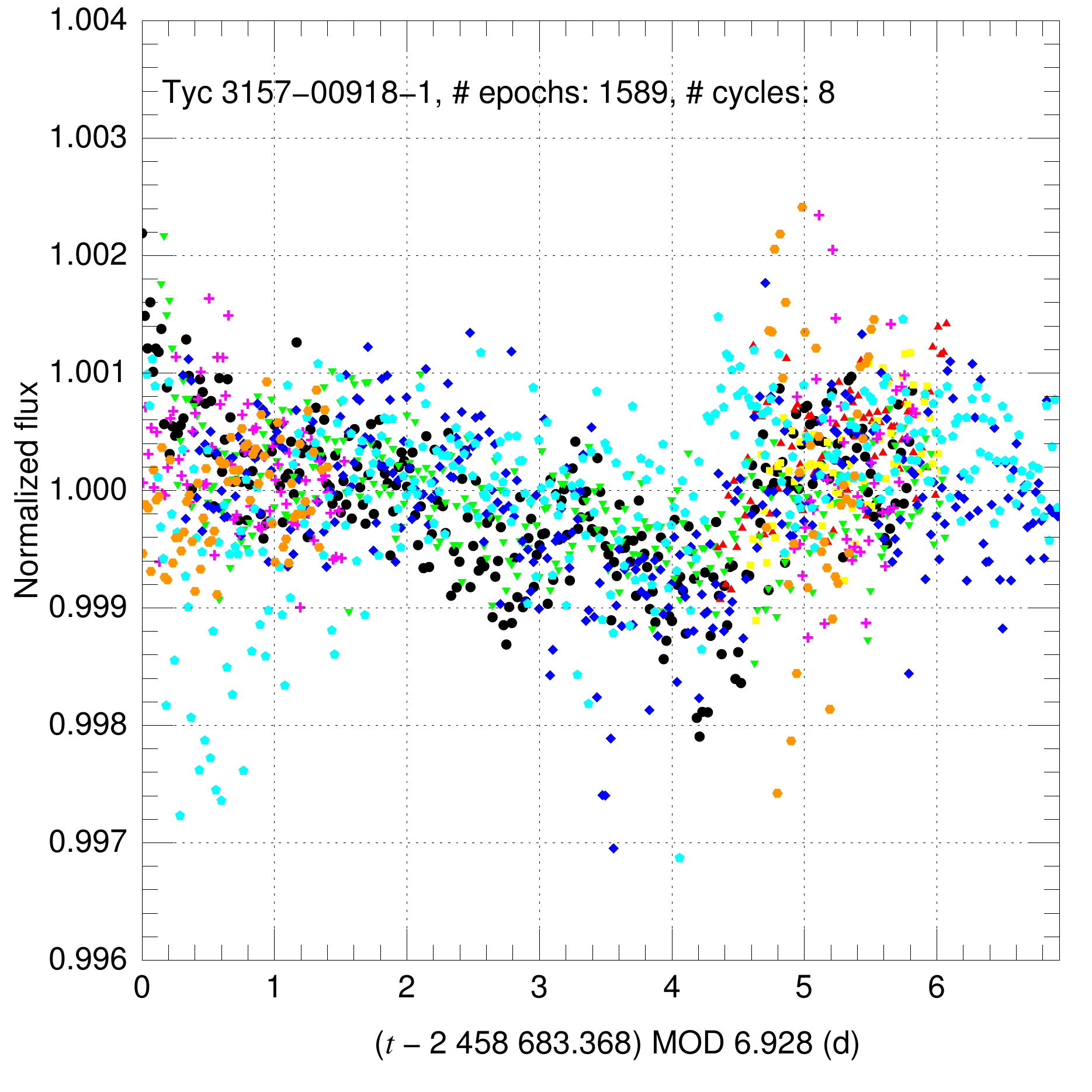} \
            \includegraphics[width=0.49\linewidth]{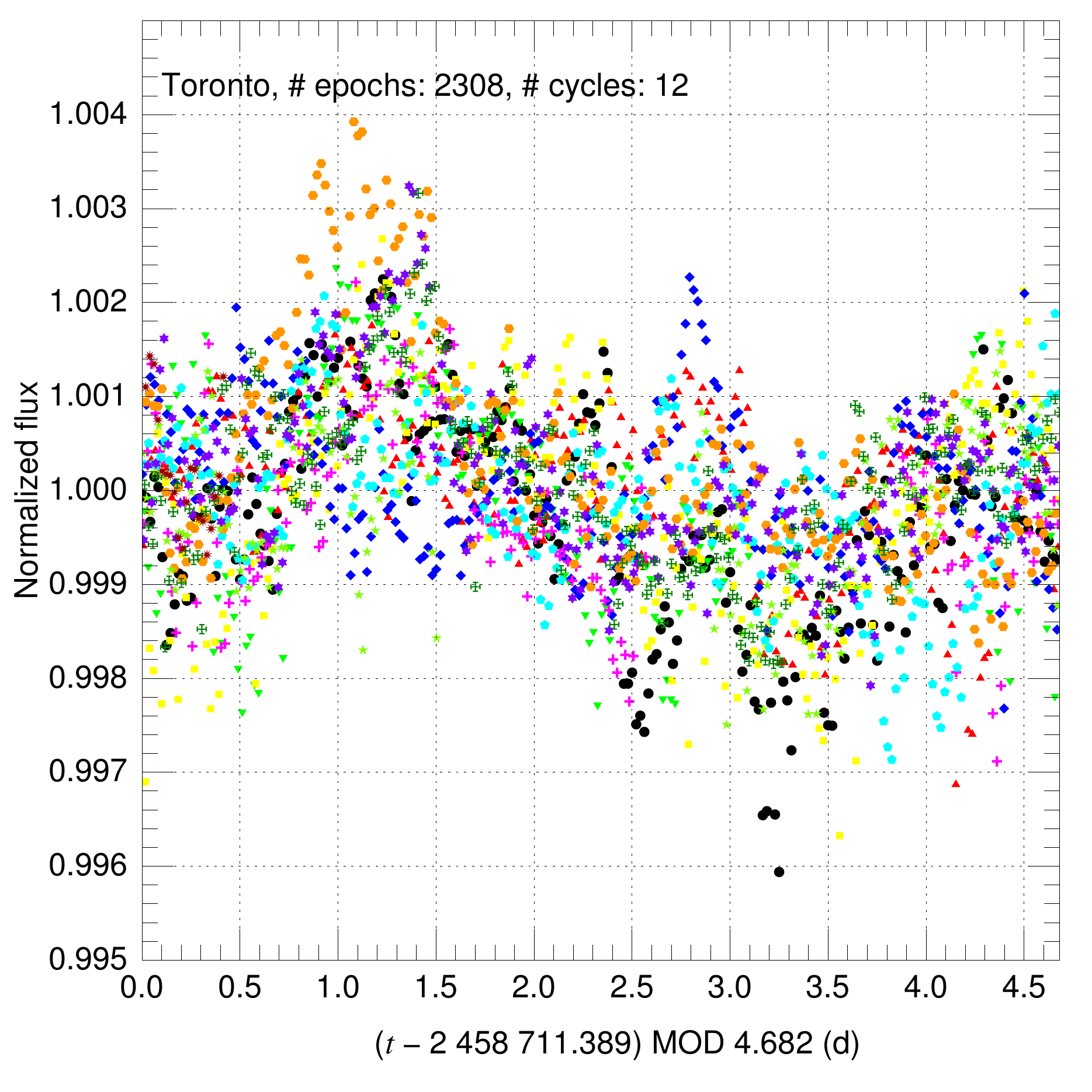}}
 \caption{Phased light curves for four of the objects analyzed with TESS. Symbols and colors are used to differentiate cycles. The number of epochs and cycles are given. 
          The HJD of the zero point and the period can be read from the labels of the $x$ axis.}
\label{TESS}
\end{figure*}

$\,\!$\indent We have used {\it TESS} to analyse the photometric variability of the nine systems involved in ejection events from the Bermuda cluster. {\it TESS} light curves were extracted 
using the {\it lightkurve} \citep{Cardetal18} {\it Python} package (version 2.11), following the same procedure developed in \citet{Trigetal21}.
The data reveals only one eclipsing system, HD~\num{200776}, plus five systems displaying pulsational or rotational variability (Bajamar, Toronto, Tyc~3179-00576-1, HD~\num{195965}, and 
Tyc~3157-00918-1). The light curves for the other three systems, namely, Tyc~3575-01514-1 (observed in sectors 15 and 16), HDE~\num{227090} (sectors 14 and 15), and HD~\num{201795} (sector 
15) display brightness stability at the $\sigma=0.5$ mmag level for the first two stars and at the $\sigma=0.8$ mmag for the last one.

HD~\num{200776} is an eclipsing binary ($P$~=~1.51299~d) with an Algol-type light curve. Based on a spectroscopic and photometric analysis of the system, \citet{Dervetal11} proposed that the 
system is a classical Algol system, where the secondary is filling the Roche lobe interacting with the primary component, i.e. it is an evolved system. They derived absolute masses for the 
components of the system of $M_1$~=~6.1~M$_\odot$ and $M_2$~=~1.2~M$_\odot$ but they did so under the assumption of a distance of 501$\pm$5~pc, which is excluded as a possibility by the
{\it Gaia}~EDR3 data. The source of the distance discrepancy is likely their assumption of a \Teff\ of 18~kK for the primary, which is too cool for a star of spectral type B1: the system
must be more distant and massive. Their mass ratio and systemic radial velocity should be correct, however, as they are derived from the analysis of the radial velocity curve. The {\it TESS}
light curve (Fig.~\ref{TESS}), observed in sectors 15 and 16, shows a similar morphology as the light curves presented by \citet{Dervetal11}, with a large difference in the depth between 
the primary and secondary eclipses and a continuous change in between them. We propose a different scenario to explain the shape of the light curve: a bloated secondary strongly irradiated 
by the hotter primary, a type of object that fits the measured mass ratio \citep{MoeDiSt15b}. This scenario makes the system compatible with an age of $\sim$2.0~Ma. In a forthcoming work, 
we will present a detailed modeling of the system.

The {\it TESS} observations of Tyc\,3179-00756-1 correspond to sector 15. The light curve shows a small amplitude ($\sim 1.5$ mmag) regular signal with $P$~=~\num{0.8715}~d (frequency, 
$f$~=~\num{1.14742}~d$^{-1}$). This type of coherent variability has been observed in main-sequence early-B type stars \citep{Bursetal20} and it corresponds to slowly pulsating 
gravity modes with frequencies below 4~d$^{-1}$.

HD~\num{195965} was observed by {\it TESS} in sectors 15 and 16. The light curve (Fig.~\ref{TESS}) shows a strong and coherent periodic signal with $P$~=~\num{0.86811}~d (frequency, 
$f$~=~\num{1.15193}~d$^{-1}$), resembling a low-frequency pulsator \citep{Bursetal20} but rotation modulation in a fast rotator cannot be discarded.
 
Tyc\,3157-00918-1 was observed in sectors 14 and 15 and presents what is likely a rotational modulation with a period $P$~=~\num{6.928}~d (Fig.~\ref{TESS}). Alternatively, if this object is 
a similar-mass early-A-type binary as suggested above, the value could correspond to one or one-half of the orbital period.

The Bajamar star was observed in sector 15. The {\it TESS} light curve shows stochastic low-frequency variability of hours scale superimposed on a small amplitude wave with $P$~=~13.38~d. 
This value could correspond to the orbital period (see Appendix~C) or it could be an artifact of the normalization procedure applied to the light curve.

{\it TESS} observed the Toronto star in sectors 15 and 16. The light curve (Fig.~\ref{TESS}) shows a very coherent signal with a period $P$~=~4.682~d, with smaller amplitude stochastic 
variability. This photometric period is very different from the spectroscopic binary period of 48.5~d \citep{Willetal01}. The projected rotational velocity is 66~km/s \citep{SimDetal17}. 
Assuming a stellar radius of 7-8~R$_\odot$ we obtain a likely rotation period of about 4.6-5.3 d. Therefore, the photometric variability can be related to a rotation modulation of the 
O~star.

\end{appendix}

\end{document}